\documentclass[journal]{IEEEtran}
\usepackage[left=1.43cm,right=1.43cm,top=1.8cm,bottom=4.21cm]{geometry}

\usepackage[algo2e]{algorithm2e} 
\usepackage{amsmath,amssymb,amsmath,amsfonts}
\usepackage{bm}
\usepackage{bbm}
\usepackage{graphicx}
\usepackage{cite}
\usepackage{epstopdf}
\usepackage{gensymb}
\usepackage{color}
\usepackage{lipsum}
\usepackage{float}
\usepackage{epstopdf}
\usepackage[mathcal]{eucal}
\usepackage{subfiles}
\usepackage[T1]{fontenc}
\usepackage{siunitx}
\usepackage{tabulary}
\usepackage{epsfig,graphics,graphicx,subfig,amssymb,amstext,amsmath,algorithm,algorithmic,wrapfig,multirow}
\usepackage{array}
\usepackage{adjustbox}
\usepackage[dvipsnames]{xcolor}
\usepackage{cite}
\usepackage{times}
\usepackage{placeins}
\usepackage{textcomp}
\usepackage[font=small]{caption}
\usepackage{flushend}
\usepackage{color, colortbl}
\usepackage{tabularx}
\usepackage{bm}
\usepackage{enumerate}
\usepackage{tikz}
\usepackage{pgfplots}
\usepackage{epstopdf}
\usetikzlibrary{shapes,arrows}
\usepackage[utf8]{inputenc}
\usepackage{mathtools}
\usepackage[utf8]{inputenc}
\usepackage{xcolor}
\usepackage{tikz}
\usetikzlibrary{chains,arrows,calc,positioning}
\usepackage{amssymb}
\usepackage{pifont}
\usepackage{soul}
\usepackage{breqn} 
\usepackage{booktabs} 
\usepackage{multirow}
\usepackage{enumitem}
\usepackage{tabulary}
\usepackage[normalem]{ulem}
\usepackage{dblfloatfix}
\usepackage{makecell}
\usepackage{lipsum}
\usepackage{amsthm}
\usepackage{comment}
\usepackage{filecontents}
\usepackage{bbm}

\newtheoremstyle{mystyle}
  {}
  {}
  {\itshape}
  {}
  {\bfseries}
  {.}
  { }
  {}
\theoremstyle{mystyle}

\newlength \figwidth
\definecolor{bittersweet}{rgb}{1.0, 0.44, 0.37}
\definecolor{glaucous}{rgb}{0.38, 0.51, 0.71}
\definecolor{gainsboro}{rgb}{0.86, 0.86, 0.86}
\definecolor{babyblueeyes}{rgb}{0.63, 0.79, 0.95}
\definecolor{silver}{rgb}{0.75, 0.75, 0.75}
\definecolor{neoncarrot}{rgb}{1.0, 0.64, 0.26}
\definecolor{Gray}{gray}{0.9}
\definecolor{LightCyan}{rgb}{0.88,1,1}
\definecolor{BackgroundLightBlue}{rgb}{0.97,0.97,1}
\definecolor{BackgroundGray}{gray}{0.98}

\newcommand{\blue}[1]{\textcolor{blue}{#1}}

\newcommand{\green}[1]{{\textcolor[rgb]{0,0.5,0}{#1}}}

\makeatletter

 \let\oldforeign@language\foreign@language
 \DeclareRobustCommand{\foreign@language}[1]{%
   \lowercase{\oldforeign@language{#1}}}

\def\nb0{{\mathbf{0}}}
\def\nb1{{\mathbf{1}}}

\def\ncalB{{\mathcal{B}}}

\def\ncalG{{\mathcal{G}}}

\def\ncalR{{\mathcal{R}}}

\def\ncalU{{\mathcal{U}}}

\def\sinr{\mathtt{SINR}}			

\def\calB{\mathcal{B}}

\makeatother

\IEEEoverridecommandlockouts
\begin{document}

\bstctlcite{IEEEexample:BSTcontrol}

\title{Cellular Network Design for UAV Corridors via Data-driven High-dimensional Bayesian Optimization}

\author{\IEEEauthorblockN{{Mohamed Benzaghta, Giovanni Geraci, David L\'{o}pez-P\'{e}rez}, and Alvaro Valcarce} 
\thanks{M.~Benzaghta is with Universitat Pompeu Fabra (UPF), Spain. G.~Geraci is with Telef\'{o}nica Research and UPF, Spain. D.~López-Pérez is with  Universitat Politècnica de Val{è}ncia, Spain. A.~Valcarce is with Nokia Bell Labs, France.
} 
\thanks{This work was supported by 
\emph{a)} HORIZON-SESAR-2023-DES-ER-02 project ANTENNAE (101167288),
\emph{b)} the Spanish Ministry of Economic Affairs and Digital Transformation and the European Union NextGenerationEU through actions CNS2023-145384, CNS2023-144333, and the UNICO 5G I+D SORUS project, 
\emph{c)} the Spanish State Research Agency through grants PID2021-123999OB-I00 and CEX2021-001195-M, 
\emph{d)} the Generalitat Valenciana, Spain, through the  CIDEGENT PlaGenT, Grant CIDEXG/2022/17, Project iTENTE, and \emph{e)} the UPF-Fractus Chair on Tech Transfer and 6G.}
\thanks{Some of the results in this paper were presented at IEEE Globecom’23 \cite{benzaghta2023designing}.}
}

\maketitle

\begin{abstract}

We address the challenge of designing cellular networks for uncrewed aerial vehicles (UAVs) corridors through a novel data-driven approach. 
We assess multiple state-of-the-art high-dimensional Bayesian optimization (HD-BO) techniques to jointly optimize the cell antenna tilts and half-power beamwidth (HPBW). We find that some of these approaches achieve over 20\,dB gains in median SINR along UAV corridors, with negligible degradation to ground user performance.
Furthermore, we explore the HD-BO's capabilities in terms of model generalization via transfer learning, where data from a previously observed scenario source is leveraged to predict the optimal solution for a new scenario target. We provide examples of scenarios where such transfer learning is successful and others where it fails. Moreover, we demonstrate that HD-BO enables multi-objective optimization, identifying optimal design trade-offs between data rates on the ground versus UAV coverage reliability. We observe that aiming to provide UAV coverage across the entire sky can lower the rates for ground users compared to setups specifically optimized for UAV corridors. Finally, we validate our approach through a case study in a real-world cellular network, where HD-BO identifies optimal and non-obvious antenna configurations that result in more than double the rates along 3D UAV corridors with negligible ground performance loss.

\end{abstract}


\section{Introduction}

\subsection{Background and Motivation}


Robust and reliable connectivity will be essential for the development of the uncrewed aerial vehicles (UAVs) ecosystem, especially in high-impact applications including delivery services and advanced urban air mobility. Private companies and government bodies are increasingly turning to mobile networks to support UAV command and control links and data payload transmission \cite{Wu2021,GerGarAza2022,FotQiaDin2019,zeng2020uav}. 
Traditional terrestrial cellular base stations (BSs) are optimized for 2D ground-level connectivity. Consequently, UAVs are often limited to receiving signals through the weaker upper antenna sidelobes, resulting in significant signal instability during flight. Additionally, when UAVs fly above buildings, they frequently face interference from line-of-sight (LoS) signals from nearby BSs \cite{GiuNikGer2024}, which degrades their signal-to-interference-plus-noise ratio (SINR) \cite{geraci2018understanding, ZenLyuZha2019}. 

The coverage and capacity of cellular networks are greatly affected by the configuration of BS antennas, where adjustments in parameters like the down-tilt angle and half-power beamwidth (HPBW) play a crucial role in optimizing signal strength and minimizing interference. Such an optimization, also known as cell shaping, is complex due to the inter-dependencies of settings across cells. Furthermore, optimizing for UAVs flying at high altitudes, requires directing some of the radiated energy upwards towards the sky, conflicting with the needs of legecy ground users (GUEs), which benefit from down-tilted cells pointing towards the ground. To achieve 3D connectivity for end-devices allowed
to fly at different heights, it may be necessary to re-engineer the cellular network originally designed for GUEs. Traditionally, cell shaping is carried out through trial and error, using radio frequency planning tools. This approach lacks scalability, calling for automated optimization techniques that leverage recent advances in data-driven models. In this paper, we propose using high-dimensional Bayesian optimization (HD-BO) to design and optimize a cellular network for 3D connectivity.


\subsection{Related Work}
 
Our research community has been dedicated to enable UAV cellular connectivity while maintaining optimal performance for ground networks. Short-term solutions, such as time and frequency separation, offer only limited scalability for managing cellular-connected UAVs effectively. While these methods can temporarily improve connectivity by allocating specific time slots or frequency bands to UAVs, they lack the capacity to handle a large and growing number of UAVs in the sky \cite{3GPP36777, NguAmoWig2018}. Additional strategies for achieving ubiquitous aerial connectivity involve dedicating specific infrastructure for aerial services, increasing network density, and incorporating non-terrestrial networks such as LEO satellites \cite{pan2023resource, KanMezLoz2021,XiaRanMez2020a,kim2022non,mozaffari2021toward,d2020analysis,garcia2019essential,benzaghta2022uav, geraci2022integrating}. These approaches aim to enhance coverage, capacity, and reliability for UAVs, supporting seamless connectivity even in areas with limited ground-based infrastructure. Although these techniques are promising, they require new deployments or signal processing enhancements and still face challenges in achieving ubiquitous connectivity for a large number of UAVs. Another line of research focus on optimizing decisions and actions on the UAV side, such as trajectory planning, aiming at maximizing UAV coverage, and/or rate, while simultaneously minimizing the impact on GUE performance \cite{esrafilian2020three, challita2018deep, de2019cellular}. Also, there exists a line of research utilizing stochastic geometry for UAV deployment optimization, which emphasizes user-centric positioning of UAVs relative to network demand \cite{matracia2023uav, galkin2019stochastic}.

The aforementioned studies assume UAV operations without spatial restrictions, requiring networks to ensure connectivity across the entire 3D space. However, akin to terrestrial vehicles and aircraft, UAVs are expected to operate within designated aerial paths or \textit{UAV corridors}, as established by authorities \cite{cherif20213d, bhuyan2021secure}. With most UAVs traveling along corridors, network operators might focus on providing reliable connectivity within these specific areas rather than across the entire sky. This vision spurred research into optimizing UAV trajectories to better align with optimal network coverage \cite{challita2018deep, bulut2018trajectory, bayerlein2021multi}. 
Nonetheless, the establishment of UAV corridors is expected to be driven by safety and logistics considerations, not communication needs, offering limited flexibility for altering UAV paths and instead underscoring the need for specialized 3D cellular service. 

Recent studies have attempted to adjust cellular network configurations to better serve UAV corridors, mainly through ad-hoc system-level optimizations of simplified setups \cite{bernabe2022optimization, bernabe2023novel, maeng2023base, chowdhury2021ensuring} 
or through fundamental theoretical analysis \cite{karimi2023optimizing, karimi2023analysis}. 
Despite these promising advancements, the community still faces the challenge of establishing a scalable optimization framework that effectively harnesses available data to enhance practical key performance indicators (KPIs), which are often mathematically complex and difficult to optimize. Our goal is to bridge this gap through a large-scale, data-driven approach designed to tackle such shortcomings.



\subsection{Approach and Contribution}

We propose a methodology based on Bayesian optimization (BO) for designing cellular BS antenna settings and provide reliable service to both GUEs and UAVs in designated aerial corridors. 
Although BO \cite{shahriari2015taking} has previously proven useful in addressing coverage/capacity tradeoffs, optimal radio resource allocation, and mobility management \cite{dreifuerst2021optimizing, eller2024differentiable, zhang2023bayesian, maggi2021bayesian, tambovskiy2022cell, maggi2023energy, tekgul2023joint, de2023towards}, it faces limitations due to the number of decision variables it can efficiently handle---typically around twenty or fewer in continuous domains \cite{frazier2018tutorial}---which effectively limits the size of a cellular network and the number of antenna parameters that can be optimized. 
In this paper, we take the first step towards employing high-dimensional BO (HD-BO) for optimizing large-scale cellular networks, thus overcoming the limitations of traditional \emph{vanilla} BO. 

To the best of our knowledge, this paper is the first to (i) apply HD-BO tools to address a practical large-scale optimization problem in cellular networks, (ii) explore model generalization within the context of transfer learning through HD-BO, and (iii) identify optimal capacity-coverage tradeoffs using multi-objective HD-BO. Additionally, our study is the first to optimize cellular connectivity along aerial corridors in real-world scenarios. Our main contributions can be summarized as follows:

\begin{itemize}[leftmargin=*]

\item
\emph{High-dimensional BO:} 
We demonstrate the shortcomings of using vanilla BO to optimize BS antenna parameters in large-scale networks. We first address this curse of dimensionality by proposing an iterative BO framework that incorporates expert knowledge into the optimization process. 
We then apply state-of-the-art HD-BO techniques such as sparse axis-aligned subspaces (SAASBO), HD-BO via variable selection (VSBO), and trust region BO (TuRBO). We illustrate the effectiveness and deficiencies of each technique from a cellular network design perspective.


\item
\emph{Transfer learning:}
Aiming at faster convergence to optimal solutions,
and aligning with the 3GPP vision on the need for data-driven model generalization \cite{3GPP38.843}, we explore the capabilities of our proposed HD-BO approach in the context of \emph{transfer learning}. 
We use transfer learning to leverage measurement outcomes from a previously performed optimization process, denoted as the \emph{scenario source}, to predict the best solution for a new optimization, termed the \emph{scenario target}.

\item
\emph{Multi-objective optimization:}
We study the effectiveness of multi-objective HD-BO in identifying optimal design trade-offs between two contrasting objectives, namely data rates on the ground versus coverage reliability along UAV corridors, and in determining a Pareto front.

\item 
\emph{Real-world case study:}
We further validate the capability of HD-BO in optimizing cell antenna parameters in a real-world scenario corresponding to a production cellular network operating in London. We employ a 3D representation of the geographical area considered and model the site-specific channel propagation through ray tracing, accounting for the actual location of the cells and their configuration.
\end{itemize}

\subsubsection*{Paper organization}
Section~\ref{sec:system_model} 
details the system model used in our study. 
Section~\ref{sec:BO} 
introduces BO and explores optimal antenna tilt design via iterative BO. 
Section~\ref{sec:HD-BO}
introduces HD-BO and tackles joint antenna tilt and half-power beamwidth optimization via HD-BO.
Section~\ref{sec:TL}
studies HD-BO model generalization in the context of transfer learning, providing both successful and unsuccessful examples.
Section~\ref{sec:MORBO}
addresses optimal 3D capacity-coverage tradeoffs via multi-objective HD-BO.
Section~\ref{sec:RD-HDBO}
presents a case study on a real-world cellular network topology with site-specific propagation channel modeling.
Section~\ref{sec:conclusion}
concludes the paper. 

\section{System Model}
\label{sec:system_model}


We model the network deployment and propagation channel following the 3GPP specifications \cite{3GPP38901,3GPP36777}. Our assumptions are detailed in the sequel and summarized in Table~\ref{table:parameters}.%
\footnote{
In Section~\ref{sec:RD-HDBO}, we present a case study on a real-world cellular network topology with site-specific propagation channel modeling.}
%


\subsection{Network Deployment}

\subsubsection*{Ground cellular network}
We consider the downlink of a cellular network, with a total of 57 BSs deployed at a height of 25\,m. BSs are deployed on a wrapped-around hexagonal layout consisting of 19 sites with a 500\,m inter-site distance (ISD). A site comprises three co-located BSs, each creating a sector (i.e., a cell) spanning a $120^{\circ}$ angle in azimuth, having a transmit power of 46\,dBm. 
Let $\ncalB$ denote the set of BSs. We set the antenna tilt $\theta^{b} \in [-20^{\circ},45^{\circ}]$ and half-power beamwidth (HPBW) $\theta^{b}_{\text{3dB}} \in [5^{\circ},70^{\circ}]$ of each BS $b\in\ncalB$ as the object of optimization, with negative and positive angles denoting down-tilts and up-tilts, respectively. 

\subsubsection*{Ground users and UAV corridors}
The network serves all user equipment (UE), i.e., both GUEs and UAVs, whose sets are denoted as $\ncalG$ and $\ncalU$, respectively. All GUEs are distributed uniformly across the entire cellular layout at a height of 1.5\,m, with an average of 10\,GUEs per sector. Unless otherwise stated, UAVs are uniformly distributed along a predefined aerial region consisting of four corridors arranged as specified in Table~\ref{table:parameters} and illustrated in Fig.~\ref{fig:UAV_Corr_Cells}, with an average of 70 UAVs per corridor. In some cases, we also consider UAVs uniformly distributed across the whole 2D area of the cellular layout and at a fixed height, so as to compare with the case of UAV corridors.

\subsection{Propagation Channel, SINR, and Achievable Rates}

\subsubsection*{Propagation Channel} 
\begin{figure}
\centering
\includegraphics[width=\figwidth]{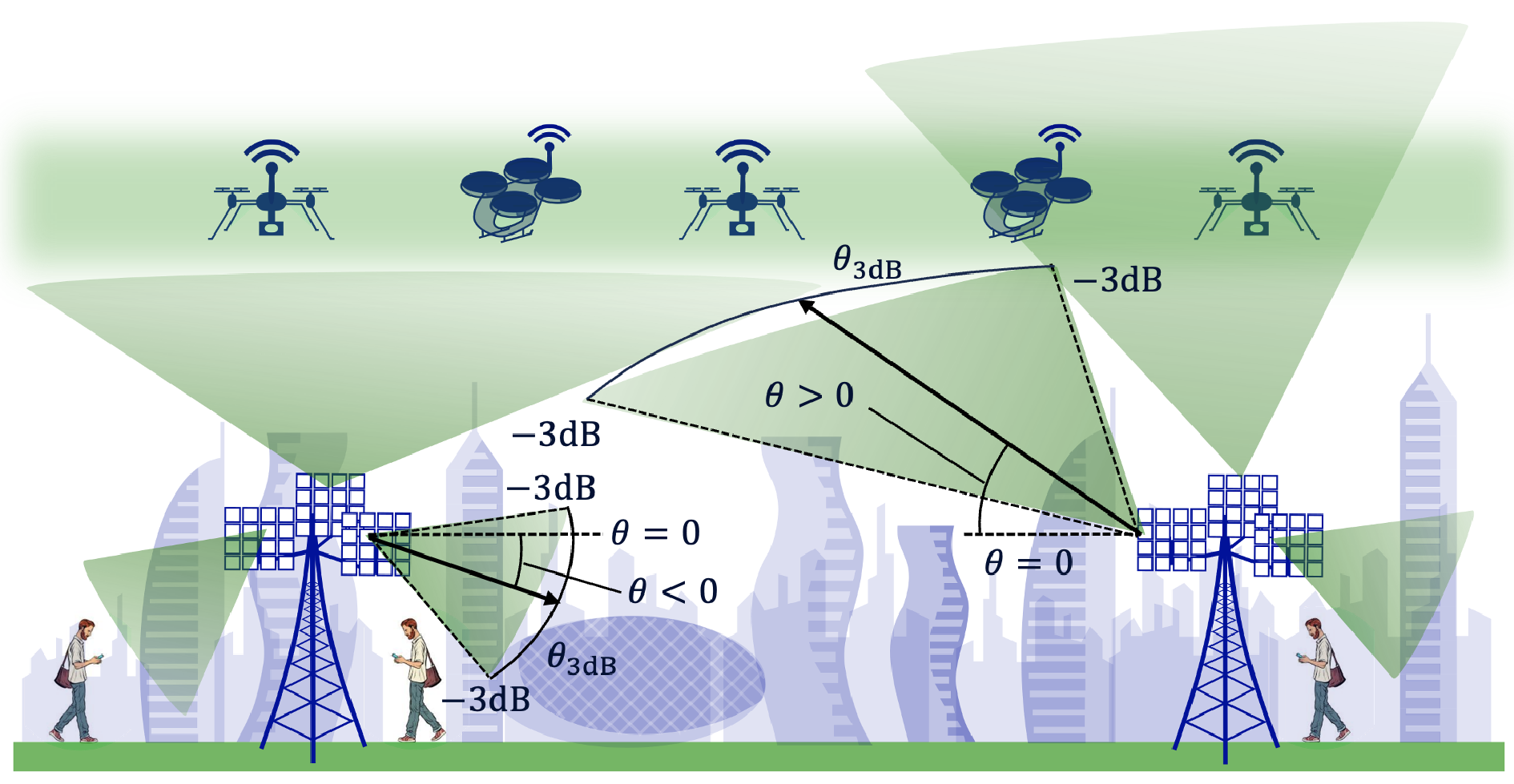}
\caption{Uptilted ($\theta>0$) and downtilted ($\theta<0$) BSs serving GUEs and UAV corridors, with $\theta$ and $\theta_{\text{3dB}}$ denoting tilt and HPBW.}
\label{fig:illustration_NEW}
\end{figure}

The network operates on a 10\,MHz band in the 2\,GHz spectrum, with the available bandwidth fully reused across all cells. All radio links experience path loss and lognormal shadow fading. We denote $G_{b,k}$ as the large-scale power gain between BS $b$ and UE $k$, which is defined as
\begin{equation}
G_{{b,k}_{\textrm{dB}}} = PL_{\textrm{dB}} + SF_{\textrm{dB}} + A(\phi, \theta)_{\textrm{dB}},
\label{eqn: LSG}
\end{equation}
where $PL_{\textrm{dB}}$ is the path loss, and $SF_{\textrm{dB}}$ is the shadow fading, defined as per 3GPP Urban Macro (UMa) channel model as per \cite{3GPP38901, 3GPP36777}  for UAVs and GUEs, respectively. $A(\phi, \theta)_{\textrm{dB}}$ is the antenna gain, depending on antenna tilt and HPBW. We characterize the BS antenna configurations by four main parameters: tilt $\theta^{b}$, bearing $\phi^{b}$, vertical HPBW $\theta^{b}_{\text{3dB}}$, and horizontal HPBW $ \phi_{\text{3dB}}$. As illustrated in Fig.~\ref{fig:illustration_NEW}, the tilt is defined as the angle between the antenna boresight and the horizon, and it can be electrically adjusted. The bearing represents the orientation of each sector. The vertical HPBW (resp. horizontal HPBW) is the angular range over which the antenna gain is above half of the maximum gain in the vertical (resp. horizontal) plane. 
Unless otherwise stated, the bearings $\phi^{b}$ and the horizontal HPBW $ \phi_{\text{3dB}}=65^{\circ}$ are assumed fixed for all BSs, and the tilt $\theta^{b}$ and HPBW $\theta^{b}_{\text{3dB}}$ are the object of optimization. 
The normalized antenna gain for a specific pair of azimuth and elevation (i.e., horizontal and vertical angles), $\phi$ and $\theta$, between a BS $b$ and a UE $k$, is \cite{3GPP38901}

\begin{equation}
A(\phi, \theta)_{\textrm{dB}} = - \min \left\{ - \left[ A_{\text{H}}(\phi) + A_{\text{V}}(\theta) \right], 25 \right\} \, ,
\label{eqn: AG}
\end{equation}
where
\begin{equation}
A_{\text{H}}(\phi)_{\textrm{dB}} = - \min \left\{ 12 \left[ \left(\phi - \phi^{b}\right)/\phi_{\text{3dB}} \right]^2, 25 \right\} \, ,
\end{equation}
\begin{equation}
A_{\text{V}}(\theta)_{\textrm{dB}} = - \min \left\{ 12 \left[ \left(\theta - \theta^{b}\right)/\theta^{b}_{\text{3dB}} \right]^2, 25 \right\} \,
\end{equation}
and where the maximum antenna gain depends on the HPBW, e.g., $\theta^{b}_{\text{3dB}}=\phi^{b}_{\text{3dB}}=65^{\circ}$ yields a maximum gain of 8\,dBi \cite{3GPP38901}.
We denote $h_{b,k}$ as the small-scale block fading between cell $b$ and UE $k$. We assume that GUEs undergo Rayleigh fading and that UAVs experience pure LoS propagation conditions, given their elevated position with respect to the clutter of buildings.%
\footnote{The small-scale fading model does not affect the conclusions drawn herein.}
Each UE $k$ is associated with the BS $b_k$ providing the largest average received signal strength (RSS). Note that the latter implies that cell association is affected by the antenna configuration, as is the case in practical systems.

\begin{table}
\centering
\caption{Network deployment and channel modeling.}
\label{table:parameters}
\def\arraystretch{1.2}
\begin{tabulary}{\columnwidth}{ |p{2.1cm} | p{5.85cm} | }
\hline
  Cellular layout				& Hexagonal grid, $\mathrm{ISD} = 500$\,m, three sectors per site, one BS per sector at $25$~m, wrap-around \\ \hline
  Frequency band 		&  $B_{b}\!=\!$ 10 MHz at 2~GHz \\ \hline
	BS max power 			& 46\,dBm over the whole bandwidth \\ \hline \hline   

  GUE distribution 				& 10 per sector on average, outdoor, at 1.5\,m \\ \hline
  	\multirow{5}{*}{UAV distribution}  &  Uniform in four aerial corridors  at 150\,m height with 2D coordinates: \\ 
	 				& $[-650,-610] \times [-780,780]$\\ 
	 				& $[-780,780] \times [-650,-610]$\\ 
	 				& $[-780,780] \times [610,650]$\\ 
	 				& $[610,650] \times [-780,780]$\\
      & 70 UAVs per corridor on average (see also Fig.~\ref{fig:UAV_Corr_Cells})\\ \hline

      UAVs/GUEs ratio 				& 50\% as per 3GPP Case~5 \cite{3GPP36777} \\ \hline
      
	User association				& Based on RSS (large-scale fading) \\ \hline

    User receiver 		& Omnidirectional antenna, 9\,dB noise figure \\ 
 \hline\hline

	Large-scale fading 		& Urban Macro as per \cite{3GPP36777, 3GPP38901} 
           \\ \hline
	Small-scale fading		& GUEs: Rayleigh. UAVs: pure LoS. \\ \hline
\end{tabulary}
\end{table}




\subsubsection*{SINR}

The downlink SINR in dB experienced by UE $k$ from its serving BS $b_k$ on a given time-frequency physical resource block (PRB) is given by
\begin{equation}
  \sinr_{\textrm{dB},k} = 10\,\log_{10} \left( \,\frac{p_{b_k} \cdot G_{b_k,k} \cdot |h_{b_k,k}|^2 }{
  \sum\limits_{b\in\calB\backslash b_k}{p_{b} \cdot G_{b,k} \cdot |h_{b,k}|^2 \,+\, \sigma_{\textrm{T}}^2}}\right),
  \label{SINR_DL_TN}
\end{equation}
where $\sigma_{\textrm{T}}^2$ denotes the thermal noise power and $p_{b}$ denotes the transmit power of BS $b$ on a PRB. It is important to note that the large-scale power gain $G_{b,k}$ implicitly depends on the antenna gain pattern, which is a function of the tilt $\theta^{b}$ and the vertical HPBW $\theta^{b}_{\text{3dB}}$. For clarity and brevity, we omit this functional dependency in (\ref{SINR_DL_TN}), while acknowledging that $G_{b,k}$ is inherently influenced by these antenna configuration parameters.

\subsubsection*{Achievable Rates}

The rate $\ncalR_k$ achievable by user $k$ served by BS $b_k$ can be related to its SINR as
\begin{equation}
    \ncalR_k = \eta_k B_k \, \mathbb{E} \left[ \log_2 (1 + \sinr_{k}) \right],
    \label{rates}
\end{equation}
where $B_k$ is the bandwidth allocated to user $k$ and $\eta_k$ the fraction of time user $k$ is scheduled by the serving BS $b_k$, thus allowing to model general radio resource allocation policies.
The expectation is taken over the small-scale fading. Without loss of generality, we assume each BS $ b\in\calB$ to multiplex its set of associated users $\ncalG_{b} \cup \ncalU_{b}$ in the time domain, 
yielding $\eta_k = \vert \ncalG_{b} \cup \ncalU_{b} \vert^{-1}$, 
and to allocate the entire available band $B_{b}$ to the scheduled user $k$, i.e., $B_k = B_{b}$.

\section{Antenna Tilt Design via Bayesian Optimization}
\label{sec:BO}


In this section, we formulate the antenna tilt optimization problem and propose a solution based on BO. Our goal is to determine the set of antenna tilts that maximize the UE rates in (\ref{rates}). The problem is defined as follows:

\begin{equation}
\max_{\bm{\theta}} \;\; \;  f_{\bm{\theta}}= \lambda \cdot \sideset{}{_{u\in\ncalU}}\sum{\log  \ncalR_u(\bm{\theta})} + (1-\lambda) \cdot \sideset{}{_{g\in\ncalG}}\sum{\log  \ncalR_g(\bm{\theta})},
\label{eqn:Opt_problem}
\end{equation}

\begin{align}
\text{s.t.} \quad & \theta_b \in \left( \underline{\theta}_b, \overline{\theta}_b \right), \enspace b = 1, \ldots, \ncalB \tag{7a}
\end{align}

\noindent where $\ncalR(\bm{\theta})$ is the achievable rate defined in (\ref{rates}) under a specific configuration of antenna tilts $\bm{\theta}$. The vector $\bm{\theta}$ = [$\theta_{1}$,\ldots,$\theta_{\calB}$] contains the antenna tilts $\theta_{b}$ of all BSs $b$ $\in$ $\calB$. The smallest allowed value is $\underline{\theta}_b$, while $\overline{\theta}_b$ is the largest allowed value. The parameter $\lambda\in[0,1]$ trades off GUE and UAV performance. As special cases, $\lambda=0$ and $\lambda=1$ optimize the cellular network for GUEs only and UAVs only, respectively. The approach in (\ref{eqn:Opt_problem}) simplifies the problem into a single-objective one. To address the trade-offs between conflicting objectives, in Section~\ref{sec:MORBO} we illustrate how Pareto front analysis can reveal the full spectrum of optimal solutions.
%

The optimization problem (\ref{eqn:Opt_problem}) is nonconvex due to the nonconcavity of the utility function $f_{\bm{\theta}}$ \cite{tekgul2023joint}. The defined joint optimization of GUEs and UAVs is challenging due to the following reasons:
\begin{itemize}
\item \textit{Conflicting Requirements:} Optimizing for UAVs operating at high altitudes necessitates redirecting a portion of the radiated energy upwards to effectively serve aerial users. However, this approach creates a challenge as it conflicts with the requirements of legacy GUEs, who rely on down-tilted cells focused on delivering coverage towards the ground.
\item \textit{Interdependence:} In a multicell environment, inter-cell interference introduces interdependence between the optimal settings of neighboring cells. The ideal antenna configuration for one cell heavily relies on the settings of its neighboring cells, creating a feedback loop that complicates finding globally optimal solutions.
\item \textit{High-Dimensionality:} The optimization problem involves a parameter for each cell, which can vary continuously within a specified range. As the number of parameters and cells increases, the combined search space grows exponentially, making exhaustive search infeasible, even with discretized search spaces. Optimization techniques such as evolutionary algorithms (e.g., PSO, NSGA-II \cite{pan2023resource}) and reinforcement learning (e.g., DDPG \cite{dreifuerst2021optimizing}) involve substantial computational expense, especially with large numbers of cells and parameters \cite{tekgul2023joint}. Additionally, in real-world cellular networks, such methods can configure antennas in ways that cause significant performance degradation. This challenge necessitates the use of computationally efficient optimization techniques, as trial-and-error approaches are prohibitively costly and time-consuming.
\end{itemize}

While our proposed optimization framework is amenable to maximize any desired key performance indicator (KPI)---including RSS, SINR, sum-rate, and any function thereof---we opted for the sum-log-rate in (\ref{eqn:Opt_problem}) since it is frequently used in cellular systems for proportional fairness and load balancing \cite{kelly1998rate}. The sum-log-rate strikes a balance between overall rate and fairness among UEs, addressing the limitations of sum-rate maximization, which tends to overlook cell size and can lead to resource allocation skewed towards a few UEs \cite{ye2013user, benzaghta2023designing}.


\subsection{Introduction to Bayesian Optimization (BO)}

BO works by iteratively constructing a probabilistic \textit{surrogate model} of the objective function $f(\cdot)$ based on prior evaluations at a number of points \cite{shahriari2015taking}. 
The surrogate model is easier to evaluate than $f(\cdot)$ and it is updated with each point evaluated. An \textit{acquisition function} $\alpha(\cdot)$ is then used to score the response from the surrogate and decide which point in the search space should be evaluated next. 
The acquisition function balances exploration (searching for new and potentially better solutions) and exploitation (focusing on the current best solutions).

\subsubsection*{Objective function evaluation}
We define a query point $\textbf{x} = \bm{\theta}$ as a configuration of the antenna tilts $\theta^{b}$ of each BS $b$ $\in$ $\calB$, and obtain the corresponding value of $f(\textbf{x})$ from (\ref{eqn:Opt_problem}).
The objective $f(\cdot)$ being optimized is a mathematically intractable stochastic function capturing the model detailed in Section~\ref{sec:system_model} and the inherent randomness of the UE locations and the statistical 3GPP channel. In this paper, we evaluate $f(\cdot)$ through system-level simulations, with each evaluation at a given point $\textbf{x}$ yielding a noisy sample $\tilde{f}(\textbf{x})$. In practice, these samples could also be obtained through real-time measurements. For convenience, let us define $\textbf{X} = [\textbf{x}_1,\ldots,\textbf{x}_N]$ as a set of $N$ query points 
and $\textbf{f}(\textbf{X})=[f_1,\ldots,f_N]^\top$ as the set of corresponding evaluations, with $f_i = f(\textbf{x}_i)$, $i=1,\ldots,N$. 

\subsubsection*{Gaussian Process prior distribution}
We use a Gaussian process (GP) prior, $\widehat{f}(\cdot)$, to create a surrogate model (i.e., the posterior) that approximates the objective function, $f(\cdot)$ \cite{shahriari2015taking}. The resulting GP model allows to predict the value of $\tilde{f}(\textbf{x})$ for a query point $\textbf{x}$ given the previous observations $\tilde{\textbf{f}}(\textbf{X})=\tilde{\textbf{f}}$ over which the model is constructed. 
Formally, 
the GP prior on the objective $\tilde{f}(\textbf{x})$ prescribes that, 
for any set of inputs $\textbf{X}$,
the corresponding objectives $\tilde{\textbf{f}}$ are jointly distributed as
\begin{equation}
  p(\,\tilde{\textbf{f}}\,) = \mathcal{N}(\,\tilde{\textbf{f}} \,\,|\,\, \boldsymbol{\mu}(\textbf{X}),\mathbf{K}(\textbf{X})\,),
  \label{posterior}
\end{equation}
where $\boldsymbol{\mu}(\textbf{X}) = [\mu(\mathrm{\textbf{x}}_1),\ldots,\mu(\mathrm{\textbf{x}}_N)]^\top$ is the $N \times 1$ mean vector, 
and $\mathbf{K}(\textbf{X})$ is the $N \times N$ covariance matrix, 
whose entry $(i,j)$ is given by the covariance $k(\textbf{x}_{i},\textbf{x}_{j})$. 
For a point $\textbf{x}$, 
the mean $\mu(\textbf{x})$ provides a prior knowledge on $f(\textbf{x})$, 
while the kernel $\mathbf{K}(\textbf{X})$ indicates the uncertainty across pairs of values of \textbf{x}. 

\subsubsection*{Gaussian Process posterior distribution}
Given a set of observed noisy samples $\tilde{\textbf{f}}$ at previously sampled points $\textbf{X}$, the posterior distribution of $\widehat{f}(\textbf{x})$ at point $\textbf{x}$ can be obtained as \cite{frazier2018tutorial} 
\begin{equation}
  p(\widehat{f}(\textbf{x}) = \widehat{f} \,\, | \,\, \textbf{X}, \textbf{$\tilde{\textbf{f}}$} \,) = \mathcal{N}(\widehat{f} \,\,|\,\, \mu(\textbf{x} \,\,|\,\, \textbf{X}, \tilde{\textbf{f}}),\sigma^2(\textbf{x} \,\,|\,\, \textbf{X}, \tilde{\textbf{f}})),
  \label{posterior_Noisy}
\end{equation}
with
\begin{equation}
  \mu(\textbf{x} \,|\, \textbf{X},\tilde{\textbf{f}}) = \mu(\textbf{x}) + \tilde{\textbf{k}}(\textbf{x})^\top (\tilde{\textbf{K}}(\textbf{X}))^{-1}(\tilde{\textbf{f}}-\boldsymbol{\mu}(\textbf{X})),
  \label{Mean_posterior_Noisy}
\end{equation}
\begin{equation}
  \sigma^2(\textbf{x} \,|\, \textbf{X},\tilde{\textbf{f}}) = k(\textbf{x},\textbf{x}) - \tilde{\textbf{k}}(\textbf{x})^\top (\tilde{\textbf{K}}(\textbf{X}))^{-1} \,\tilde{\textbf{k}}(\textbf{x}),
  \label{Kernel_posterior_Noisy}
\end{equation}
where 
$\tilde{\textbf{k}}(\textbf{x}) = [k(\textbf{x},\textbf{x}_{1}),\ldots,k(\textbf{x},\textbf{x}_{N})]^\top$ is the $N \times 1$ covariance vector and $\tilde{\textbf{K}}(\textbf{X}) = \textbf{K}(\textbf{X}) + \sigma^2 \textbf{I}_{\text{N}}$, with $\sigma^2$ denoting the observation noise represented by the variance of the Gaussian distribution, and $\textbf{I}_{\text{N}}$ denoting the $N \times N$ identity matrix.
Note that \eqref{Mean_posterior_Noisy} and \eqref{Kernel_posterior_Noisy} represent the mean and variance of the estimation $\widehat{f}(\textbf{x})$, the latter indicating the degree of confidence.  


\subsection{Antenna Tilt Optimization via Iterative BO}

We now propose a novel algorithm to iteratively optimize each of the BS antenna tilts in $\bm{\theta}$ one by one. We denote this approach as \emph{iterative-BO} to distinguish it from other approaches that jointly optimize all antenna tilts.%
\footnote{We omit the pseudo-code of our iterative-BO algorithm due to space limitations. We provide an example for max-SINR optimization in \cite[Algorithm~1]{benzaghta2023designing}.}

\subsubsection*{Initial dataset creation}
The proposed iterative-BO algorithm starts by creating a GP prior $\{\mu(\cdot), k(\cdot, \cdot)\}$ based on a dataset $\mathcal{D} = \{\textbf{x}_1,\ldots,\textbf{x}_{N_{\textrm{o}}},\tilde{f}_1,\ldots,\tilde{f}_{N_{\textrm{o}}}\}$ containing $N_{\textrm{o}}$ initial observations. The dataset is constructed via system-level simulations according to the objective function  defined in \eqref{eqn:Opt_problem} and the model detailed in Section~\ref{sec:system_model}. The antenna tilts $\bm{\theta}_i$ for every observation point $\textbf{x}_i = \bm{\theta}_i$ in $\mathcal{D}$ are chosen randomly in $[-20^{\circ}, 45^{\circ}]$.

\subsubsection*{Iterative approach}
Once the initial GP prior is constructed, the vector $\bm{\theta}_0$ is initialized with all entries set to $0^{\circ}$. We denote $\tilde{f}^{*}$ as the best observed objective value, which is initialized to $\tilde{f}_0^{*} = -\infty$. 
The algorithm then iterates over each BS $b \in \calB$, one at a time. At every iteration, the BS considered is 
\begin{equation}
b_{n} = ((n-1) \mod \|\ncalB\|) + 1,
\end{equation}
and for each such iteration $n$, only the antenna tilt of the BS $b_{n}$ under consideration is updated, while keeping the remaining entries of $\bm{\theta}_n$ fixed to their values from the previous iteration. The query point under optimization is thus reduced to a scalar that we denote as $\widehat{x}_{n} = \theta_{b_{n}}$. 

\subsubsection*{Acquisition function}
The algorithm then leverages the observations in $\mathcal{D}$ to choose $\widehat{x}_{n}$. This is performed via an acquisition function $\alpha(\cdot)$, which is designed to trade off the exploration of new points in less favorable regions of the search space with the exploitation of well-performing ones. 
The former prevents getting caught in local maxima, 
whereas the latter minimizes the risk of testing points with excessively degrading performance. We adopt the expected improvement (EI) as the acquisition function, which has shown to perform well in terms of balancing the trade-off between exploration and exploitation \cite{shahriari2015taking, maggi2021bayesian}. At iteration $n$, the EI tests and scores a set of $N_{\textrm{c}}$ randomly drawn candidate points $\{\widehat{x}_{\text{cand}_{1}},\ldots,\widehat{x}_{\text{cand}_{N_{\textrm{c}}}}\}$ through the surrogate model, i.e., the posterior (\ref{posterior_Noisy}). The EI is defined as \cite{huang2006sequential,maggi2021bayesian} 
\begin{equation}
    \begin{aligned}
    \alpha \left(\widehat{x}_{\text{cand}} \,|\, \mathcal{D} \right) =\, & [\, \mu\,(\widehat{x}_{\text{cand}} \,|\, \mathcal{D})\, - \widehat{f}^{*} - \xi] \cdot \Phi(\delta)\\
    & +\sigma^2\,(\widehat{x}_{\text{cand}} \,|\, \mathcal{D}) \cdot \phi(\delta),
    \label{eqn:EI}
    \end{aligned}
\end{equation}
where $\widehat{f}^{*} = 
{\textrm{max}_i}\,\{\widehat{f}_{\text{cand}_{i}}\}$ denotes the current best approximated objective value according to the surrogate model, $\Phi$ (resp. $\phi$) is the standard Gaussian cumulative (resp. density) distribution function, and
\begin{equation}
    \delta = \frac{\mu\,(\widehat{x}_{\text{cand}} \,|\, \mathcal{D})\, - \widehat{f}^{*} - \xi}{\sigma^2\,(\widehat{x}_{\text{cand}} \,|\, \mathcal{D})\,},
    \label{eqn:EI_delta}
\end{equation}
with $\mu(\widehat{x}_{\text{cand}}\,|\, \mathcal{D})$ and $\sigma^2\,(\widehat{x}_{\text{cand}} \,|\, \mathcal{D})\,$ given in \eqref{Mean_posterior_Noisy} and \eqref{Kernel_posterior_Noisy}, respectively. 
The parameter $\xi\in[0,1)$ in (\ref{eqn:EI}) and (\ref{eqn:EI_delta}) regulates exploration vs. exploitation, with larger values promoting the former, and vice versa. In this paper, we aim for a risk-sensitive EI acquisition function and set $\xi = 0.01$ \cite{maggi2021bayesian}. 

\subsubsection*{Batch evaluation of candidate points}
Given the safety and operational constraints inherent in real-world UAV network connectivity, we adopt a batch evaluation strategy to efficiently explore the query space while minimizing the number of physical experiments.
At each iteration, a set of $N_{\textrm{c}}=500$ candidate points is selected based on the posterior distribution (\ref{posterior_Noisy}) and evaluated in parallel across available computational resources.
This approach leverages the ability of Bayesian optimization to learn from limited samples, making it particularly well-suited for problems where extensive real-world experimentation with UAVs is impractical or risky.
We split the candidate points into 10 batches each consisting of 50 points. The query
point $\widehat{x}_{n}$ is then chosen as
\begin{equation}
    \widehat{x}_{n} = \underset{\substack{i}}{\textrm{arg}\,\textrm{max}}\;\; \alpha \left(\widehat{x}_{\text{cand}_i} \,|\, \mathcal{D} \right).
    \label{eqn:BO_acq}
\end{equation} 

\subsubsection*{Objective function evaluation}
Once $\widehat{x}_{n} = \theta_{b_{n}}$ is determined, the vector $\bm{\theta}_n$ is obtained from $\bm{\theta}_{n-1}$ by replacing its $b_{n}$-th entry with $\theta_{b_{n}}$, yielding $\textbf{x}_n = \bm{\theta}_n$. A new observation of the objective function $\tilde{f}(\textbf{x}_n)$ is then produced, and the dataset $\mathcal{D}$, the GP prior, and the best observed objective value $\tilde{f}^{*}$ are all updated. 

The algorithm then moves on to optimizing the antenna tilt of BS $b_{n+1}$, until all BSs in $\calB$ have been optimized. 
This loop over all BSs is then repeated until the best observed value $\tilde{f}^{*}$ has remained unchanged for a certain number of consecutive loops, $\ell_{\text{max}}$, after which the algorithm recommends the point $\textbf{x}^{*}$ that produced the best observation $\tilde{f}^{*}$.


\subsection{Convergence and Performance of Iterative-BO}

We now compare the performance of iterative-BO with two benchmarks, indicated as the 3GPP baseline (where all BSs are down-tilted) and the benchmark vanilla BO (i.e., the standard BO algorithm, in which all the variables under optimization are considered at the same time). The experiments are performed for the value of $\lambda$ = 0.5, which correspond to optimizing the cellular network for GUEs and UAVs with equal weight. 
Iterative-BO is run on BoTorch, an open-source library built upon PyTorch \cite{balandat2020botorch}. 
We use the Matern-5/2 kernel for $\mathbf{K}(\textbf{X})$ and fit the GP hyperparameters using maximum posterior estimation.

\subsubsection*{Performance of iterative-BO}

\begin{figure}
\centering
\includegraphics[width=\figwidth]{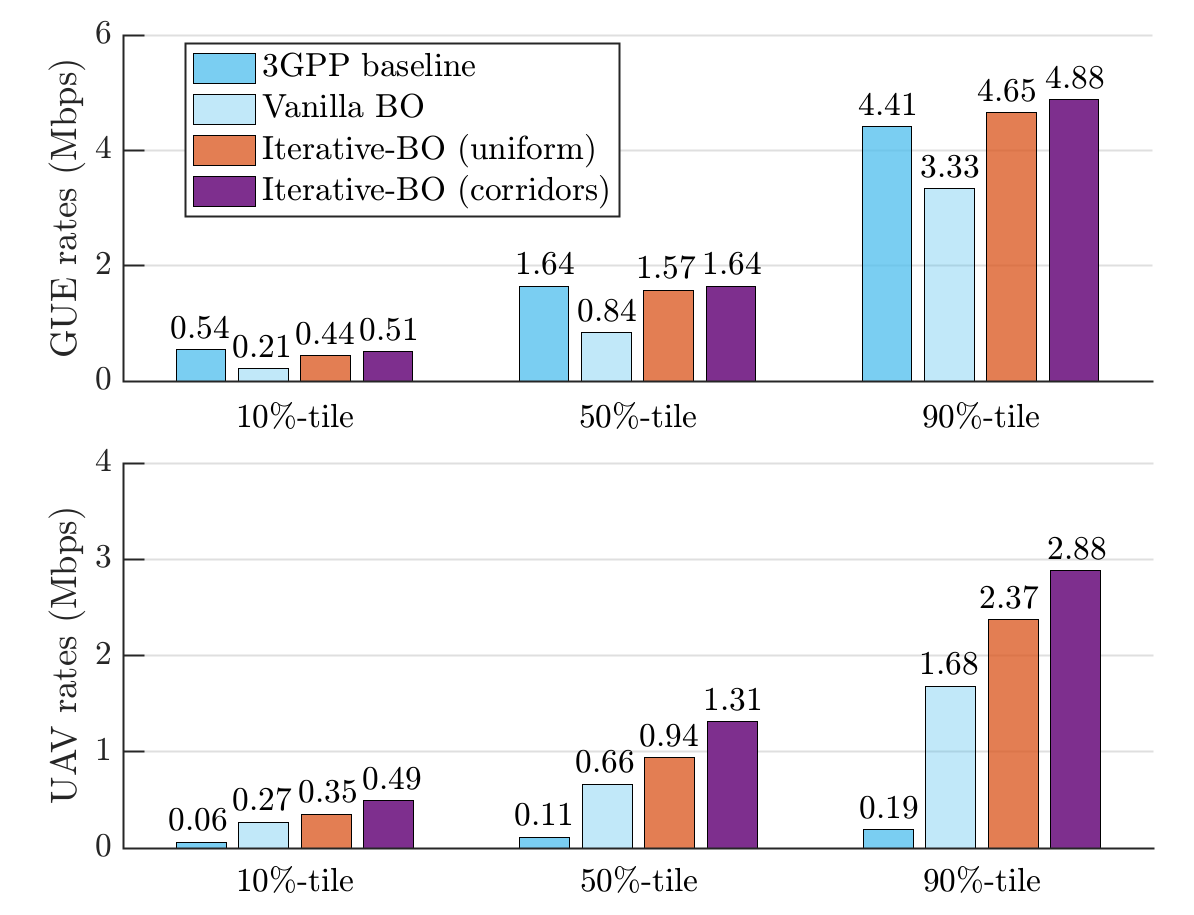}
\caption{Performance comparison of the iterative-BO framework with the benchmark 3GPP baseline and benchmark vanilla-BO, indicating the achievable rates of GUEs (top) and UAVs rates (bottom).}
\label{fig:Opt_vanilla_Rate_compare}
\end{figure}

We implement iterative-BO in two scenarios: (i) A scenario where the UAVs are uniformly distributed across the cells, at an altitude of 150\,m and with an average density of 5 UAVs per cell (`uniform'); (ii) The corridor deployment detailed in Table~\ref{table:parameters} (`corridor'). Fig.~\ref{fig:Opt_vanilla_Rate_compare} shows a comparison of the achieved rates of both GUEs and UAVs:
\begin{itemize}[leftmargin=*]
\item 
With a traditional, all-downtilt 3GPP baseline configuration (blue), where all BSs are down-tilted to $-12^{\circ}$ with all vertical HPBW (vHPBW) set to $10^{\circ}$ \cite{3GPP38901}, all UAVs experience rates at least one order of magnitude lower than their GUE counterpart in median. Similar performance is observed for both scenarios (i.e., UAVs uniformly distributed and in corridor deployment).
\item
Antenna tilt optimization through iterative-BO significantly improves the UAV rates, with 8$\times$ and 12$\times$ gains in median for the cases of uniformly distributed UAVs (red) and UAVs on corridors (purple), respectively, with respect to the all-downtilt 3GPP baseline (blue). 
\item 
Such UAV performance improvement does not correspond to a severe GUE performance degradation, with the median GUE rate only reduced by less than 5\% in the case of uniform UAVs and preserved in the case of UAV corridors.
\item 
When compared to vanilla BO (light blue), for corridor deployment, iterative-BO  (purple) enhances the GUE rates by 142\%, 95\%, and 46\% for the 10\%-tile, 50\%-tile, and 90\%-tile, respectively. Moreover, it increases the UAVs rates by 80\%, 98\%, and 70\% for the 10\%-tile, 50\%-tile, and 90\%-tile, respectively. Note that vanilla BO performs the same for UAVs uniformly distributed and in corridor deployment.
\end{itemize}

\subsubsection*{Convergence of iterative-BO}

\begin{figure}
\centering
\includegraphics[width=\figwidth]{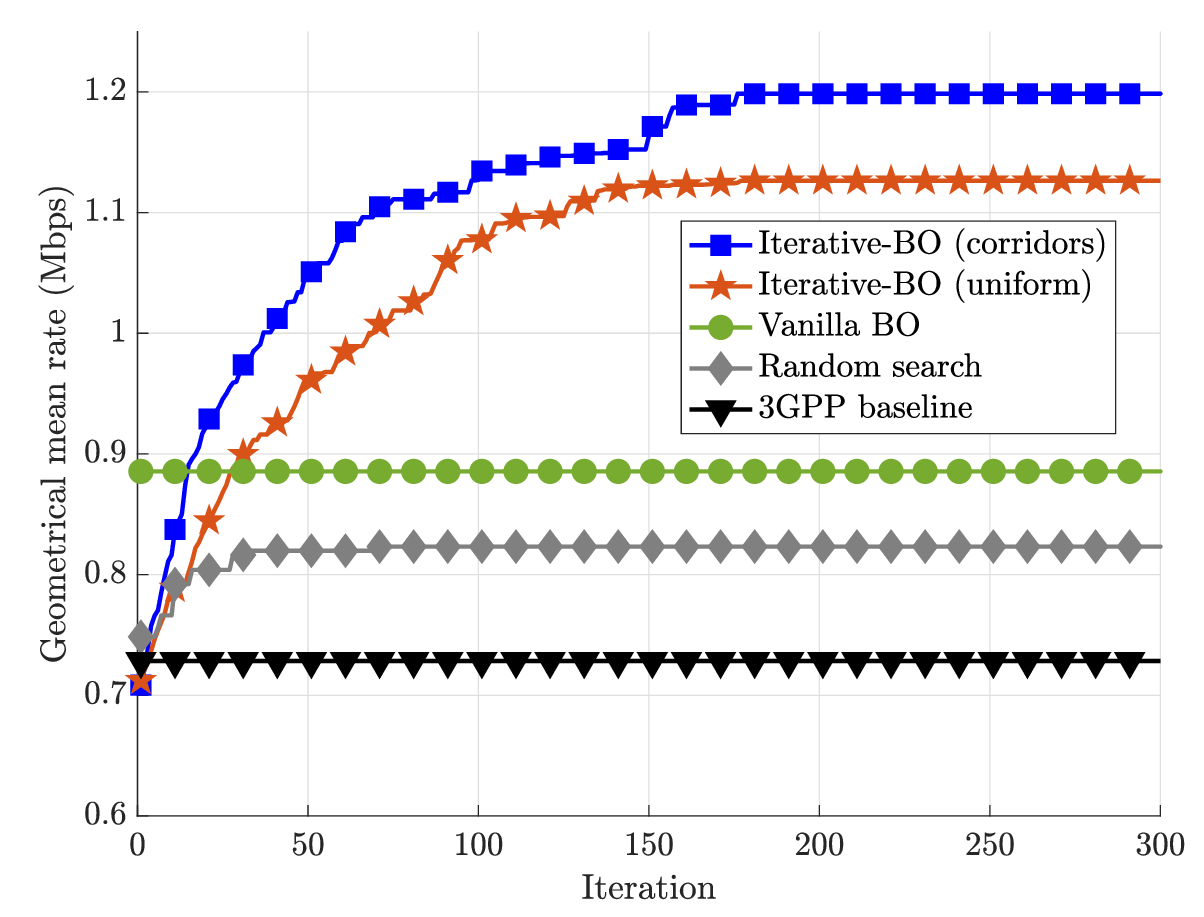}
\caption{Best observed objective function vs. number of iterations $n$.}
\label{fig:Opt_vanilla_conv_compare}
\end{figure}

Fig.~\ref{fig:Opt_vanilla_conv_compare} compares the performance of vanilla BO to the proposed iterative-BO, by illustrating the best observed objective value at each iteration $n$. To show a quantity of practical interest, rather than the function $f(\cdot)$ defined in (\ref{eqn:Opt_problem}), we plot the geometrical mean of the UEs rate:
\begin{equation}
\overline{\mathcal{R}} = \left( \sideset{}{_{k\in\ncalU\cup\ncalG}}\prod \ncalR_k \right)^{\frac{1}{|\ncalU\cup\ncalG|}},
\end{equation}
which for $\lambda=0.5$ can be readily mapped to $f(\cdot)$ as 
\begin{equation}
\overline{R} = e^{{f(\cdot)}/{|\ncalU\cup\ncalG|}}.
\end{equation}
Two other benchmarks are also shown, namely `random search' and all-downtilt `3GPP baseline', where tilts are sampled at random in $[-20^{\circ}, 45^{\circ}]$ and uniformly set to $-12^{\circ}$, respectively. 
Fig.~\ref{fig:Opt_vanilla_conv_compare} shows that vanilla BO struggles with high-dimensional settings, such as optimizing antenna tilts for 57 cells. 
The proposed iterative-BO addresses these limitations, achieving convergence (i.e., a stable solution) in about 150 iterations. For the case of UAV corridors, iterative-BO improves the geometrical mean rate by more than 30\% and 60\% with respect to vanilla BO and the 3GPP baseline, respectively.

\subsubsection*{Optimal antenna tilts}

\begin{figure}
\centering
\includegraphics[width=\figwidth]{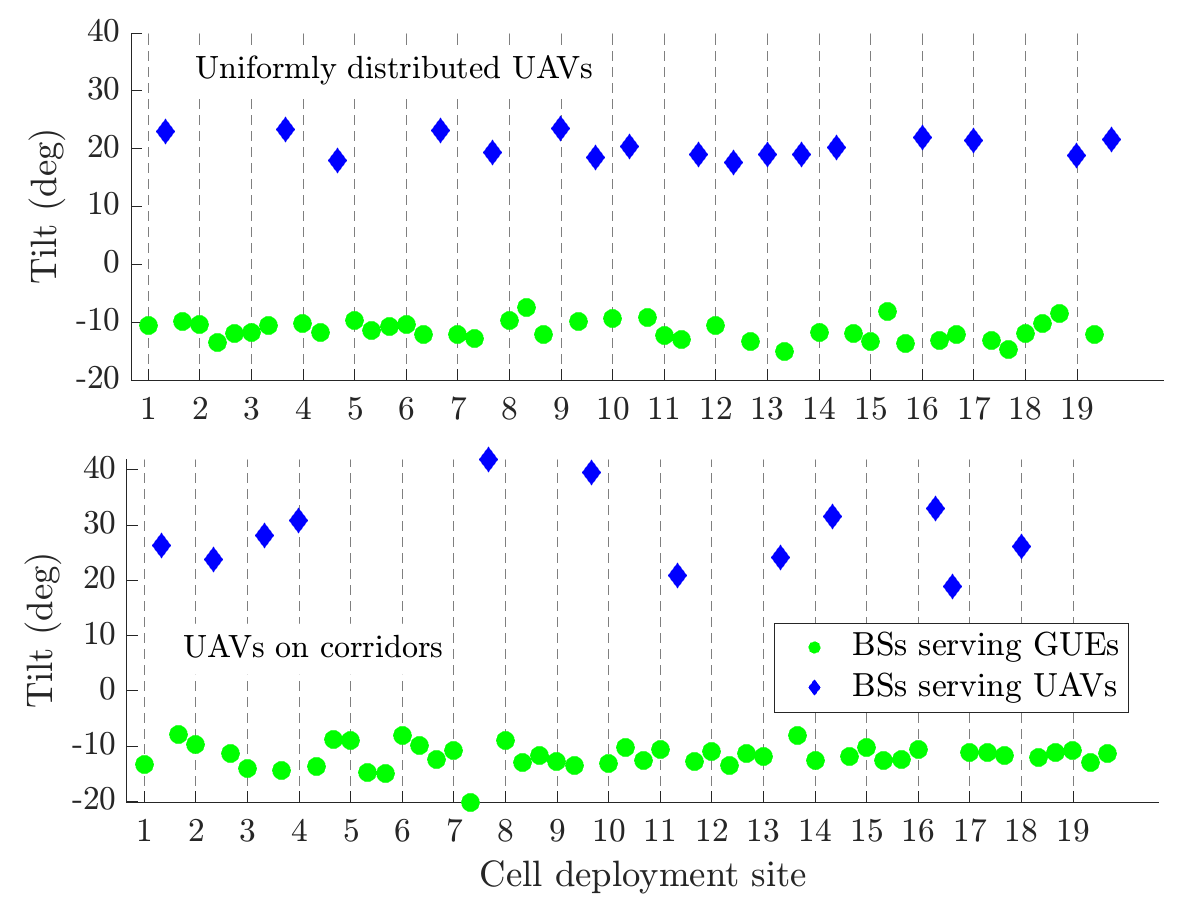}
\caption{Optimized tilts for $\lambda = 0.5$ when UAVs are uniformly distributed (top) and confined to corridors (bottom). Green circles and blue diamonds denote BSs serving GUEs and UAVs, respectively.}
\label{fig:Opt_values_tilt}
\end{figure}

Fig.~\ref{fig:Opt_values_tilt} shows the optimal values found by our iterative-BO algorithm for the electrical antenna tilts $\bm{\theta}$. Each numbered deployment site comprises three cells and can be identified through the black dots in Fig.~\ref{fig:UAV_Corr_Cells}. Markers indicate whether each cell is serving GUEs (green circles) or UAVs (blue diamonds). Unlike a traditional cellular network configuration where all BSs are down-tilted \cite{3GPP38901}---e.g., the all-downtilt 3GPP baseline---pursuing a trade-off between GUEs and UAVs results in a subset of the BSs being up-tilted with the rest remaining down-tilted. Such configuration is non-obvious and would be difficult to design heuristically. Specifically, the optimum configuration results in up-tilting 12 BSs for the case of corridors, compared to 17 for the uniform scenario. As a result, a better performance is achieved in the case of UAVs confined to corridors as the network requires fewer cells to cover the sky.
%
Fig.~\ref{fig:UAV_Corr_Cells} and Fig.~\ref{fig:corrmat_comp_a} show the resulting cell partitioning of the UAV corridors and of the ground, respectively.

{\begin{figure}[!t]
    \centering
    \subfloat[Cell partitioning at 150\,m height.]{
        \includegraphics[width=0.48\figwidth]{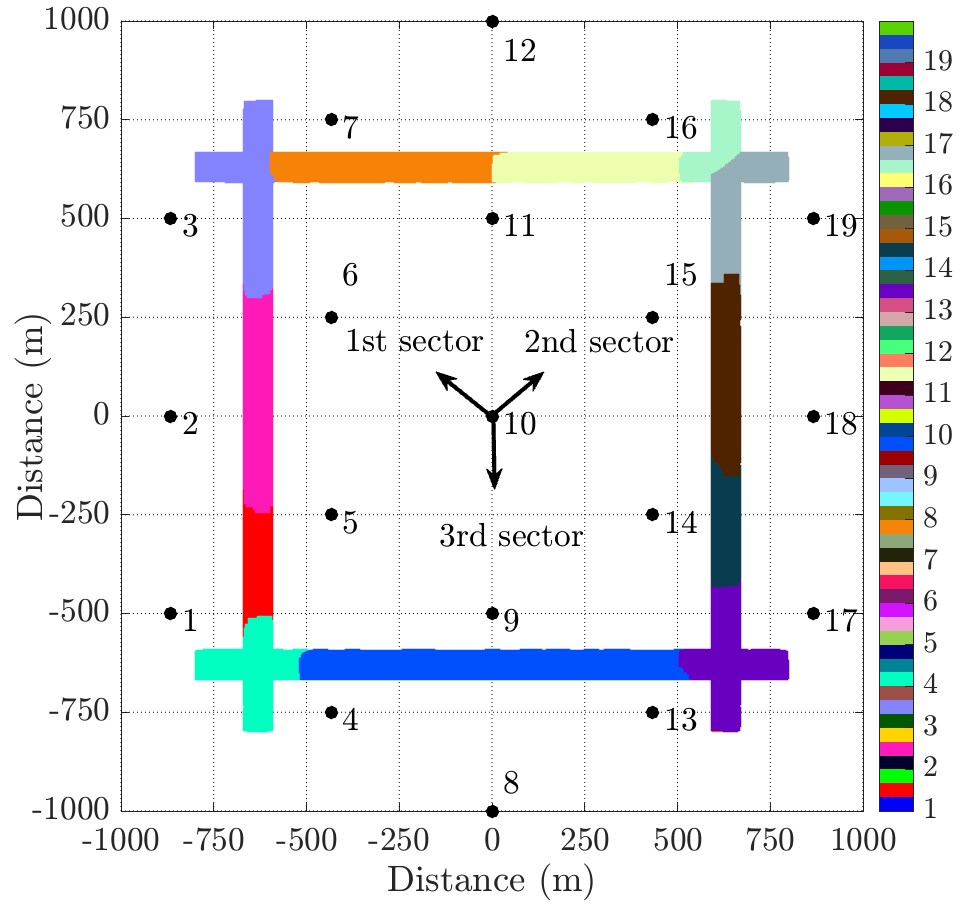}
        \label{fig:UAV_Corr_Cells}
    }
    \hspace{0.3mm}
    \subfloat[Cell partitioning at 1.5\,m height.]{
        \includegraphics[width=0.48\figwidth]{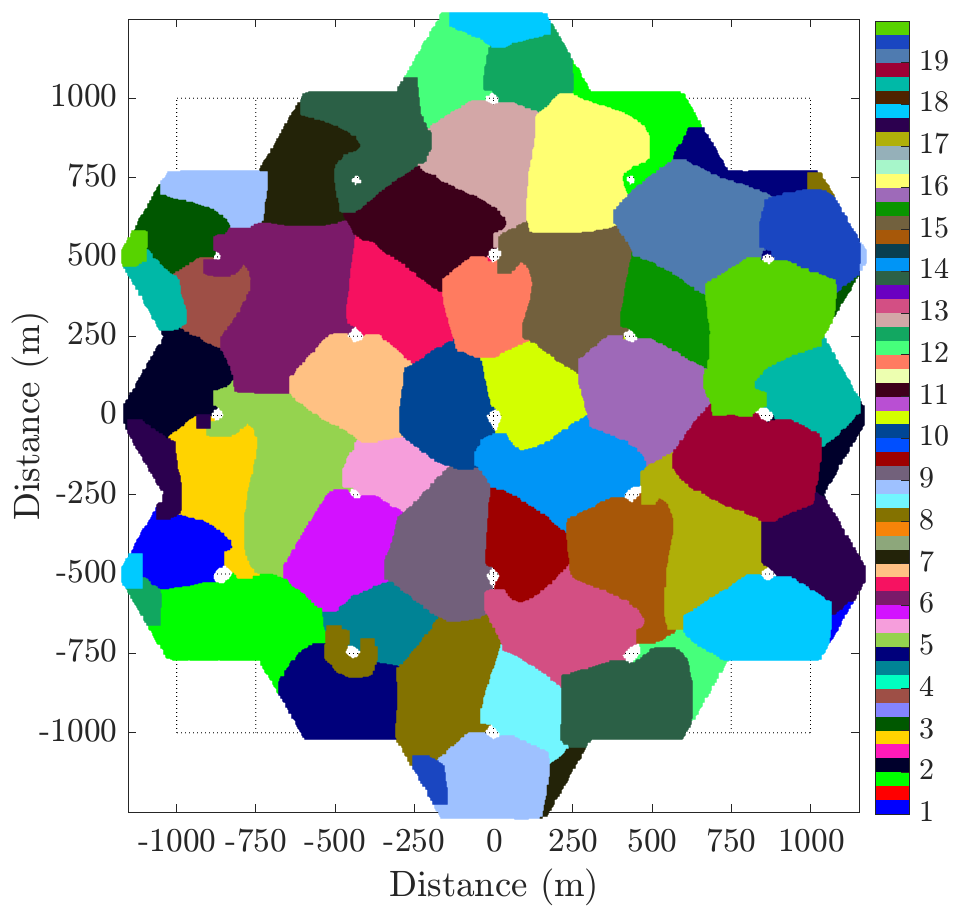}
        \label{fig:corrmat_comp_a}
    }
    \caption{Cell partitioning (a) along UAV corridors and (b) on the ground when BS antennas are optimized for both GUEs and UAVs ($\lambda$ = 0.5).}
    \label{fig:GUE_Corr_Cells}
\end{figure}}

\subsubsection*{Drawbacks of iterative-BO}
While iterative-BO overcomes the limitations of vanilla BO in handling a large number of optimization variables, the GP prior it creates is not generalized to the whole search space, but rather a local prior specific to the BS tilt being optimized at each iteration. To overcome this issue, in Section~\ref{sec:HD-BO} we employ HD-BO leveraging a general prior. This then allows us to apply transfer learning in Section~\ref{sec:TL} and to carry out multi-objective optimization in Section~\ref{sec:MORBO}.


\section{Joint Antenna Tilt and Half-power Beamwidth Design via High-dimensional Bayesian Optimization}
\label{sec:HD-BO}

For BO methods to be more sample-efficient, it is necessary to introduce a hierarchical significance for the dimensions of $\textbf{x} \in D$. For instance, in a high-dimensional problem, features $\{\textbf{x}_2, \textbf{x}_{47}\}$ may play a crucial role in capturing the main variation of the objective function $f$, while features $\{\textbf{x}_5, \textbf{x}_{32}, \textbf{x}_{112}\}$ may have a moderate significance. The rest of the features may be of negligible importance. These conditions are exploited in high-dimensional BO (HD-BO). In the following, we introduce three state-of-the-art HD-BO methods and evaluate their performance in (i) solving problem (\ref{eqn:Opt_problem}) and (ii) solving an even higher-dimensional problem: the joint antenna tilt and HPBW optimization defined in (\ref{eqn:Opt_problem_joint}).
%
%

\subsection{Introduction to High-dimensional BO (HD-BO)}


In the sequel, we introduce the core features of three HD-BO methods: Sparse Axis-Aligned Subspaces (SAASBO), BO via Variable Selection (VSBO), and Trust Region BO (TuRBO). For further technical details, we refer the reader to 
\cite[Section~4]{eriksson2021high}, \cite[Section~3]{shen2021computationally}, and \cite[Section~2]{eriksson2019scalable}, respectively.

\subsubsection*{Sparse Axis-Aligned Subspaces (SAASBO)}

The authors in \cite{eriksson2021high} introduced a prior model that focuses on sparse axis-aligned subspaces, prioritizing feature relevance hierarchically and incorporating a broad spectrum of smooth non-linear functions for efficient (approximate) inference. This approach uses a sparse structure Gaussian process (GP) model, characterized by (inverse squared) length scales $\rho_i$, defined as follows:
\begin{equation}
\begin{aligned}
\text{[kernel variance]} \quad & \sigma^2_k \sim \mathcal{LN}(0, 10^2) \\
\text{[global shrinkage]} \quad & \tau \sim \mathcal{HC}(\alpha) \\
\text{[length scales]} \quad & \rho_i \sim \mathcal{HC}(\tau) \quad \text{for } i = 1, \ldots, D. \\
\text{[function values]} \quad & \textbf{f} \sim \mathcal{N}(0, K^{\psi}_{XX}) \quad \text{with } \psi = \{\rho_1:d, \sigma^2_k\} \\
\end{aligned}
\end{equation}
where $\mathcal{LN}$ is the log-Normal distribution, $\mathcal{HC}$ the half-Cauchy distribution, and $\alpha > 0$ a hyperparameter adjusting shrinkage intensity. In the SAAS framework, global shrinkage (i.e., sparsity) is controlled by $\tau > 0$, which tends towards zero due to the half-Cauchy prior. The inverse squared length scales $\rho_i$ follow half-Cauchy priors, concentrating near zero and effectively deactivating most dimensions. If the observations provide sufficient evidence, the posterior over $\tau$ shifts towards higher values, reducing shrinkage and allowing more $\rho_i$ to deviate from zero, thus `activating' additional dimensions. As more data is gathered, more $\rho_i$ depart from zero, guiding the posterior towards a more relevant set of dimensions. This mechanism contrasts with traditional GP models fitted via maximum likelihood estimation, which tend to overfit in high-dimensional settings.


\subsubsection*{BO via variable selection (VSBO)}

The BO via variable selection (VSBO) framework, introduced in \cite{shen2021computationally}, addresses the optimization of a black-box function $f(\textbf{x}) \colon \mathcal{X} \to \mathbb{R}$ within the domain $\mathcal{X} = [0,1]^D$ for large dimensions $D$. VSBO operates under the premise that the variables in $\textbf{x}$ can be divided into important variables $\textbf{x}_{\text{ipt}}$ and unimportant variables $\textbf{x}_{\text{nipt}}$. Every $N_{vs}$ iterations, VSBO updates these sets, focusing only on $\textbf{x}_{\text{ipt}}$ during each BO iteration to fit the GP model. A new set of important variables is then determined by maximizing the acquisition function. VSBO constructs a conditional distribution $p(\textbf{x}_{\text{nipt}} \mid \textbf{x}_{\text{ipt}}, \mathcal{D})$ based on an initial dataset $\mathcal{D}$, from which it selects values for $\textbf{x}_{\text{nipt}}$ that are predicted to maximize $f(\textbf{x})$ when $\textbf{x}_{\text{ipt}}$ is fixed. Variable importance is quantified using a gradient-based importance score (IS) method, termed Grad-IS, which assesses the average magnitude of the partial derivatives of $f$ with respect to each variable; a large average derivative suggests a high importance. As the direct derivative of $f$ is typically unknown, VSBO estimates the expected gradient of the GP's posterior mean, normalized by the posterior standard deviation. Variables are sequentially selected based on their IS. When a new variable is added, the GP model is refitted using the currently selected variables, and a new final loss is calculated. If this new loss closely matches the previous loss---computed excluding the new variable---the selection process concludes, confirming the importance of the selected variables. Importantly, VSBO learns these axis-aligned subspaces automatically, without requiring any pre-defined hyperparameters. 


\subsubsection*{Trust Region BO (TuRBO)}

To address the challenges of high dimensionality in BO, the authors in \cite{eriksson2019scalable} proposed Trust Region BO (TuRBO), an approach that shifts from global surrogate modeling to managing multiple independent local models. Each model focuses on a separate region of the search space. TuRBO achieves global optimization by simultaneously operating several local models and allocating samples using an implicit multi-armed bandit strategy, enhancing the acquisition strategy's effectiveness by directing samples to promising local optimization efforts. TuRBO utilizes trust region (TR) methods from stochastic optimization, which are gradient-free and employ a simple surrogate model within a defined TR---typically a sphere or polytope centered around the best solution found. However, simple surrogate models may require overly small trust regions for accurate modeling. To address this, TuRBO employs a GP surrogate model within the TR, preserving global BO features such as noise robustness and systematic uncertainty handling. In TuRBO, the TR is defined as a hyperrectangle centered at the optimal current solution, $f^{*}$. The side length of the TR is initialized to $L \gets L_{\text{init}}$. Each dimension's side length is then adjusted according to its respective length scale $\lambda_i$ in the GP model, while maintaing a total volume of $L^{d}$. The side length for each dimension is given by
\begin{equation}
L_i = {\lambda_i L}\cdot{\left(\sideset{}{_{j=1}^d}\prod \lambda_j \right)^{-1/d}}\!\!\!\!\!\!.
\end{equation}

During each local optimization run, an acquisition function selects a batch of $q$ candidates at each iteration, ensuring they remain within the designated TR. If the TR's side length $L$ were large enough to cover the entire search space, this method would be equivalent to standard global BO. Thus, adjusting $L$ is crucial: the TR needs to be large enough to encompass good solutions but compact to ensure the local model's accuracy. The TR is dynamically resized based on optimization progress: it is doubled ($L \gets \min \{L_{\text{max}}, 2L\}$) after $\tau_{\text{succ}}$ consecutive successes, and halved ($L \gets L/2$) after $\tau_{\text{fail}}$ consecutive failures. Success and failure counters are reset after adjustments. If $L$ falls below $L_{\text{min}}$, that TR is discarded and a new one is initiated at $L_{\text{init}}$. The TR's side length is capped at $L_{\text{max}}$. TuRBO maintains $m$ trust regions simultaneously: $\text{TR}_{l}$, $l \in \{1, \dots, m\}$, each a hyperrectangle of base side length $L_{l} \leq L_{\text{max}}$. Candidate selection involves choosing a batch of $q$ candidates from the union of all trust regions. Thompson sampling is used for selecting candidates within and across TRs.


\subsection{Joint Antenna Tilt and HPBW Optimization via HD-BO}

We now present the results from applying the three HD-BO frameworks---SAASBO (Benchmark \#1), VSBO (Benchmark \#2), and TuRBO---to a joint antenna tilt and HPBW optimization problem, detailing their effectiveness and shortcomings.
on the following two problems:
\begin{itemize}[leftmargin=*]
\item The antenna tilt optimization problem (\ref{eqn:Opt_problem}).
\item A joint antenna tilt and HPBW optimization problem.%
\end{itemize}

\noindent Indeed, besides antenna tilts, coverage and capacity in a cellular network are affected by the antenna half-power beamwidth (HPBW), which can be optimized to enhance received signal strength and minimize interference. 
Adding the vector of vertical HPBW $\bm{\theta}_{\text{3dB}}$ as an additional optimization parameter to the defined problem formulation in (\ref{eqn:Opt_problem}) yields:

\begin{align}
\max_{\bm{\theta},\bm{\theta}_{\text{3dB}}} \;\; \;  f_{\bm{\theta},\bm{\theta}_{\text{3dB}}} & = \lambda \cdot \sideset{}{_{u\in\ncalU}}\sum{\log  \ncalR_u(\bm{\theta,\bm{\theta}_{\text{3dB}} })} \nonumber \\
& \quad + \; (1-\lambda) \cdot \sideset{}{_{g\in\ncalG}}\sum{\log  \ncalR_g(\bm{\theta},\bm{\theta}_{\text{3dB}}}),
\label{eqn:Opt_problem_joint}
\end{align}

\begin{align}
\text{s.t.} \quad & \theta_b \in \left( \underline{\theta}_b, \overline{\theta}_b \right), \enspace b = 1, \ldots, \ncalB \tag{20a} \\
& \theta_{{\text{3dB}}_{b}} \in \left( \underline{\theta_{\text{3dB}}}_b, \overline{\theta_{\text{3dB}}}_b \right), \enspace b = 1, \ldots, \ncalB \tag{20b}
\end{align}

\noindent where $\ncalR(\bm{\theta},\bm{\theta}_{\text{3dB}})$ is the achievable rate defined in (\ref{rates}) under a specific configuration of antenna tilts $\bm{\theta}$ and HPBWs $\bm{\theta}_{\text{3dB}}$. The vectors $\bm{\theta}$ and $\bm{\theta_{\text{3dB}}}$ contains the antenna tilts $\theta_{b}$ and HPBWs $\theta_{{\text{3dB}}_b}$, of all BSs $b$ $\in$ $\calB$, respectively. The smallest allowed values are $\underline{\theta}_b$ and $\underline{\theta_{\text{3dB}}}_{b}$, while $\overline{\theta}_b$, $\overline{\theta_{\text{3dB}}}_{b}$ are the largest allowed values. There exists a coupling between $\bm{\theta}$ and $\bm{\theta}_{\text{3dB}}$ as narrow beams offer higher gain but require precise tilt alignment, while wider beams trade gain for coverage. This interdependence makes the optimal configuration difficult to characterize analytically. Our HD-BO framework addresses this by jointly exploring the search space in a data-driven way, capturing such interactions without explicit modeling. The parameter $\lambda\in[0,1]$ trades off GUE and UAV performance. 

For our scenario with 57 cells, (\ref{eqn:Opt_problem_joint}) involves 114 optimization variables, requiring efficient methods to handle the large search space. We use $\lambda$ = 0.5, balancing the optimization for GUEs and UAVs. 
Both SAASBO and VSBO are run on BoTorch \cite{balandat2020botorch}, using Matern 5/2 as the kernel function. In SAASBO, the shrinkage hyperparameter is set to $\alpha = 0.1$. 
TuRBO is run using an open-source repository \cite{eriksson2019scalable} with the following hyperparameters: $\tau_{\text{succ}} = 3, \tau_{\text{fail}} = 15, L_{\text{init}} = 0.8, L_{\text{min}} = 2^{-7}, L_{\text{max}} = 1.6$. The domain is scaled to the unit hypercube $[0,1]^d$.
%


\subsubsection*{Convergence of HD-BO}

\begin{figure}
\centering
\includegraphics[width=\figwidth]{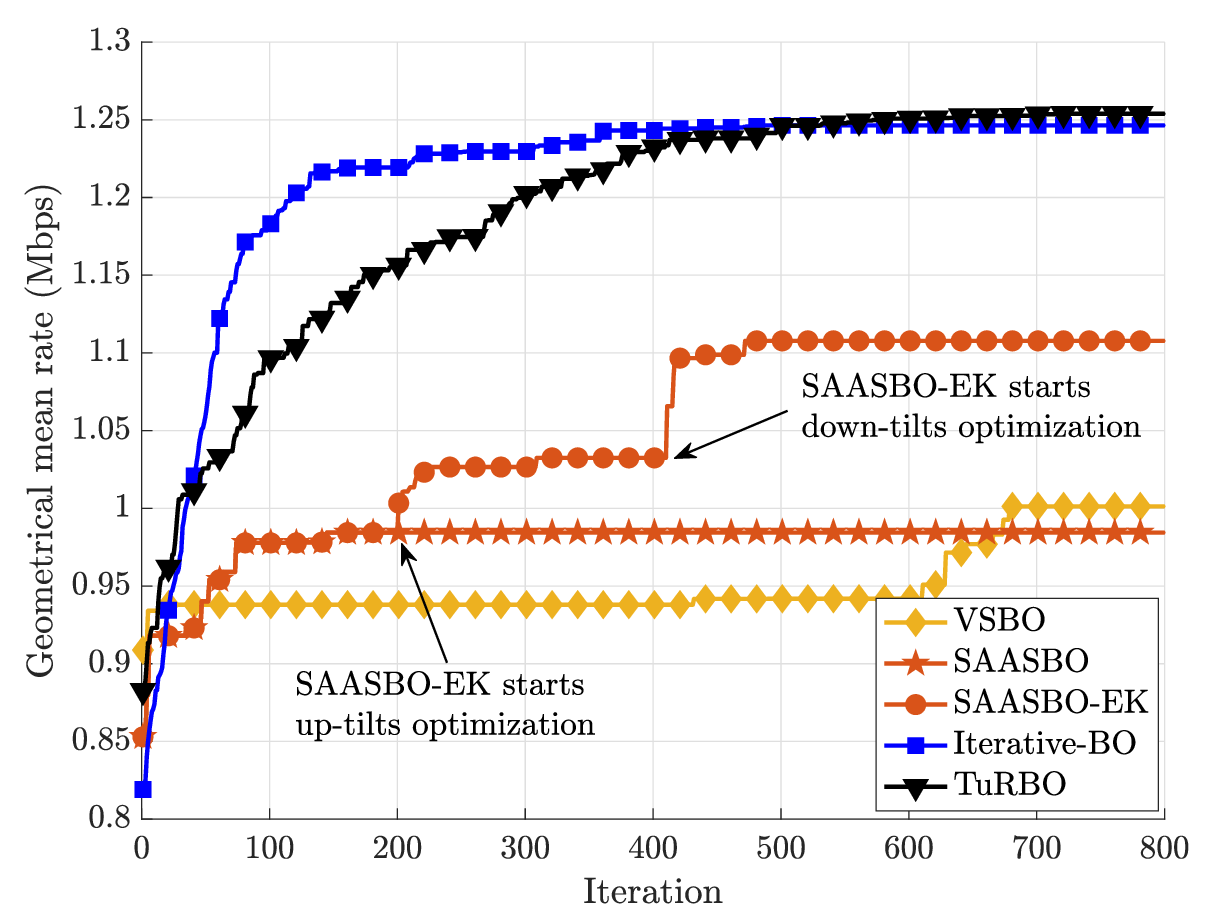}
\caption{Convergence of the proposed algorithms compared to Benchmark \#1 (VSBO) and Benchmark \#2 (SAASBO), showing the evolution of the best observed objective vs. the number of iterations $n$.}
\label{fig:Opt_HDBO_EK_compare}
\end{figure}

Fig.~\ref{fig:Opt_HDBO_EK_compare} shows the convergence of the proposed Iterative-BO, SAASBO-EK, and TuRBO compared to Benchmark \#1 (VSBO) and Benchmark \#2 (SAASBO). The figure illustrates the best observed objective at each iteration $n$. This is compared to the iterative-BO framework presented in Section~\ref{sec:BO}, when this is applied to a joint optimization of antenna tilts and HPBW. This figure leads to the following conclusions:
\begin{itemize}[leftmargin=*]
\item
VSBO performs poorly is our scenario. All cells are equally important for the problem defined in (\ref{eqn:Opt_problem}), making VSBO equivalent to a vanilla-BO approach, which struggles with more than 20 variables \cite{shen2021computationally}. 
\item 
SAASBO performs better than VSBO but not as well as iterative-BO. The difficulty lies in indicating sparsity within dimensions when optimizing both down-tilts and up-tilts simultaneously. 
\item 
The performance of SAASBO improves with expert knowledge (SAASBO-EK), indicating which cells should be up-tilted or down-tilted, optimizing each set separately until convergence. This approach gets closer to the performance of iterative-BO but lacks generalization to the whole search space, limiting its use to transfer learning or multi-objective optimization.
\item 
TuRBO performs best by effectively selecting the trust regions that impact the KPI, identifying down-tilt and up-tilt regions without expert knowledge. TuRBO creates a generalized prior, enabling transfer learning and multi-objective optimization as discussed in Sections~\ref{sec:TL} and \ref{sec:MORBO}.
\end{itemize}


\subsubsection*{Computation complexity of HD-BO} 

To understand the computational complexity of the HD-BO framework, we measured and compared the runtime of Iterative-BO, VSBO, SAASBO, and TuRBO. 
Table~\ref{tab:HDBO_runtime} presents the average runtime per iteration for the joint antenna tilt and HPBW optimization problem, relative to the runtime of Iterative-BO, with the dimensional space $D = 114$ and each method initializing with $2 D$ points.
Run times are obtained using a 2.5GHz, Intel Core i5-10300H CPU, with 16GB of RAM.
VSBO improves efficiency by only using important variables for running BO, thus reducing the runtime for both fitting the GP model and maximizing the acquisition function. Assuming that the number of variables is $p$ and the Quasi-Newton method (QN) is applied for both fitting the GP and optimizing the acquisition function, the computational complexity for each QN iteration is $\mathcal{O}(p^2 + pn^2 + n^3)$. Given that the method used for fitting the GP and maximizing the acquisition function under BoTorch is limited-memory BFGS, a QN method \cite{balandat2020botorch}, the complexity for maximizing the acquisition function is $\mathcal{O}(p^2 + pn + n^2)$, where $n$ is the number of queries already made \cite{shen2021computationally}. Since complexity is proportional to $p^2$, selecting a smaller subset of variables (thus reducing $p$) can reduce the runtime.


\begin{table}[ht]
\centering
\begin{tabular}{lcccc}
\hline
\textbf{Method} & Iterative-BO & VSBO & SAASBO & TuRBO \\ \hline
\textbf{Runtime/iteration} & 1 & $0.06$ & 2.40 & 1.40 \\\hline
\end{tabular}
\caption{Average runtime per iteration for the joint antenna tilt and HPBW optimization problem, for different HD-BO methods relative to the runtime of Iterative-BO. The dimensional space is $D = 114$ and each method initializes with $2 D$ points.}
\label{tab:HDBO_runtime}
\end{table}

\subsubsection*{Performance of HD-BO}
Due to its demonstrated higher suitability for the problem under consideration, our discussion will focus on the performance of TuRBO.
Table~\ref{tab:HDBO_performance} compares the performance of TuRBO to the 3GPP baseline, vanilla BO, and iterative-BO, by illustrating the best observed objective value (the geometrical mean rate of all UEs), along with the number of iterations $n$ needed to reach convergence. Note that no convergence values are provided for the 3GPP baseline and vanilla BO, as the former is a fixed configuration and the latter immediately gets trapped in a local optimum (see Fig.~\ref{fig:Opt_vanilla_conv_compare}).

\begin{table}[ht]
\centering
\caption{Performance comparison of TuRBO to the 3GPP baseline, vanilla BO, and iterative-BO in terms of best observed KPI value (geometrical mean rate) and convergence (number of iterations).}
{\scriptsize
\begin{tabular}{l c c}
\hline
 & \textbf{Geometrical mean rate (Mbps)} & \textbf{Convergence (\(n\))} \\
\hline
\textbf{3GPP baseline} & 0.73 & -- \\
\textbf{Vanilla BO (Tilts)} & 0.89 & -- \\
\textbf{Iterative-BO (Tilts)} & 1.20 & 176 \\
\textbf{TuRBO (Tilts)} & 1.20 & 215 \\
\textbf{TuRBO (Tilts + vHPBW)} & 1.25 & 502 \\
\hline
\end{tabular}
}
\label{tab:HDBO_performance}
\end{table}

\begin{figure}
\centering
\includegraphics[width=\figwidth]{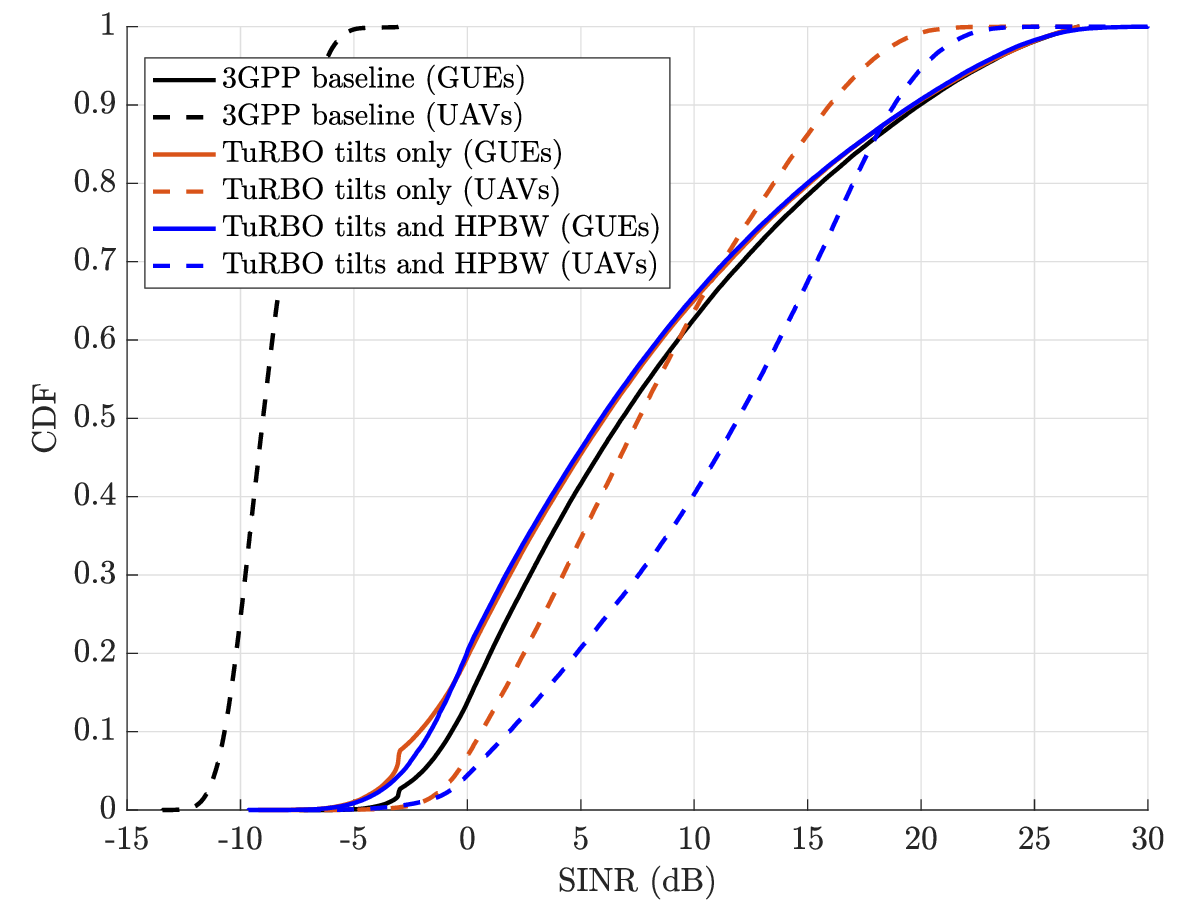}
\caption{SINR for UAVs (dashed) and GUEs
(solid) when the network is optimized for both ($\lambda$ = 0.5) and for an all-downtilt 3GPP baseline.}
\label{fig:SINR_CDFs_tilts_vHPBW}
\end{figure}

Fig.~\ref{fig:SINR_CDFs_tilts_vHPBW} shows the cumulative distribution function (CDF) of the SINR perceived by GUEs (solid lines) and UAVs (dashed lines) when the cellular network is optimized for both UEs categories ($\lambda = 0.5$). The figure shows the performance of TuRBO when optimizing only the antenna tilts (red curves) with all HPBW set to $10^{\circ}$ and when jointly optimizing antenna tilts and HPBW (blue curves). The performance of an all-downtilt 3GPP baseline configuration (black) is also shown as a baseline for comparison. The figure shows the following:
\begin{itemize}[leftmargin=*]
\item 
For the 3GPP baseline configuration, all UAVs experience SINRs below -5\,dB and can thus be regarded as being in outage (dashed black).%
\footnote{While (\ref{rates}) is based on Shannon rates, practical adaptive modulation and coding schemes require a minimum SINR, typically around -5\,dB \cite{geraci2018understanding}.}
This behavior is expected and due to a large number of line-of-sight interfering signals of comparable strength to the signal from the serving cell \cite{GerGarAza2022}. 
\item
Antenna tilt optimization through TuRBO significantly improves the UAV SINR, with median gains of around 17\,dB (dashed red). This SINR boost also reduces the percentage of UAVs in outage from 100\% to zero.
\item
The UAV performance improvement does not correspond to a severe GUE performance degradation, with the GUE SINR reduced by less than 1\,dB in median (solid black vs. solid red). 
\item
Optimizing for both antenna tilts and HPBW through TuRBO further improves UAV performance, with gains of about 21\,dB in median SINR compared to the 3GPP baseline (dashed blue vs. dashed black), again without degrading the GUE SINR.
\end{itemize}

\subsubsection*{Optimal antenna tilts and HPBW}
\begin{figure}
\centering
\includegraphics[width=\figwidth]{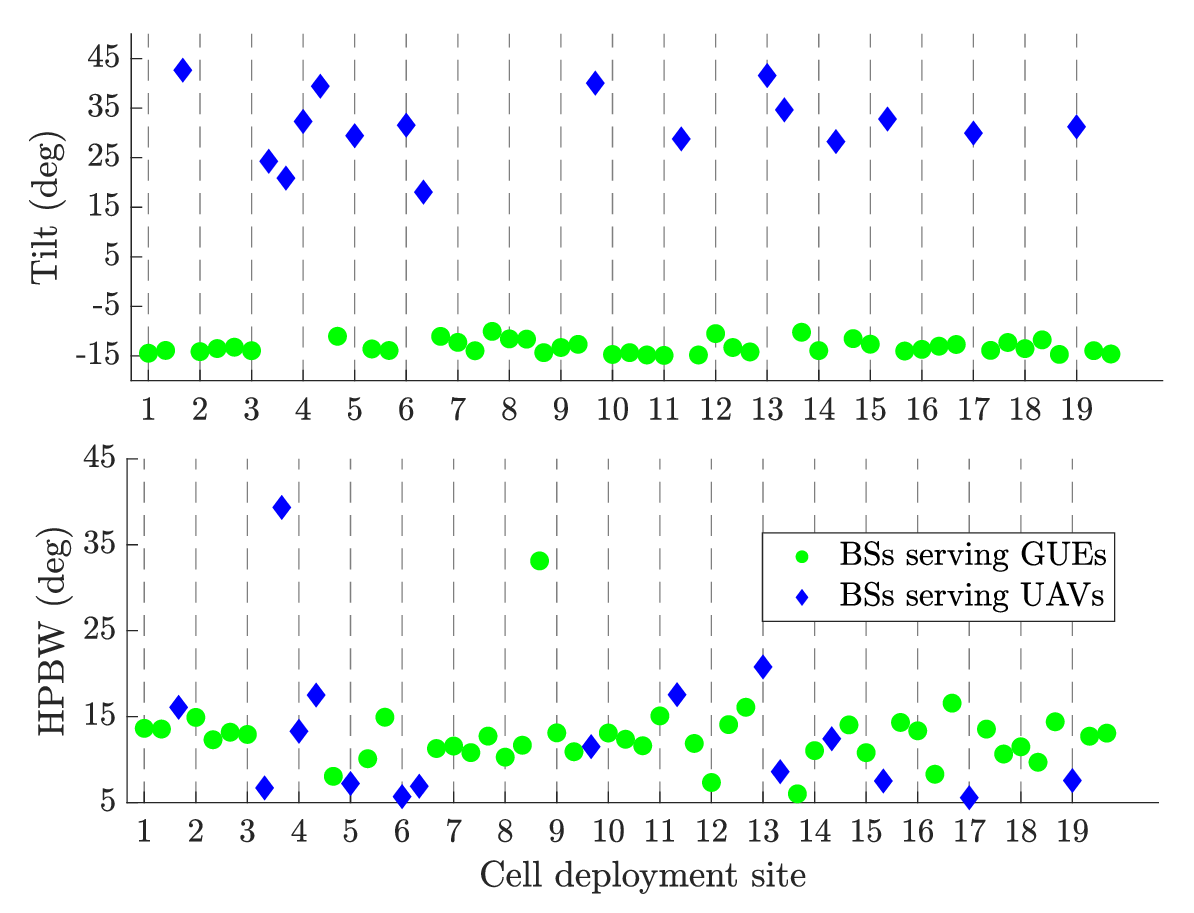}
\caption{Optimized tilts and HPBW for $\lambda = 0.5$. Green circles and blue diamonds denote BSs serving GUEs and UAVs, respectively.}
\label{fig:Opt_Tilts_vHPBW_config_TuRBO}
\end{figure}

Fig.~\ref{fig:Opt_Tilts_vHPBW_config_TuRBO} shows the optimal values of the antenna tilts $\bm{\theta}$ and HPBW $\bm{\theta}_{\text{3dB}}$ produced by TuRBO for the case $\lambda=0.5$. The index on the x-axis denotes the deployment site (black dots in Fig.~\ref{fig:UAV_Corr_Cells}), each comprising three cells. Markers indicate whether each cell is serving GUEs (green circles) or UAVs (blue diamonds). Unlike a traditional cellular network configuration where all BSs are down-tilted to $-12^{\circ}$ with a fixed vertical HPBW of $10^{\circ}$ \cite{3GPP38901}, the results show that the optimal configuration to support GUEs and UAV corridors is more complex and would be difficult to design heuristically.

\section{Transfer Learning with High-dimensional Bayesian Optimization}
\label{sec:TL}

Deploying machine learning models in commercial networks requires confidence in their ability to perform consistently across diverse scenarios, configurations, and site-specific conditions. This necessity emphasizes the importance of understanding a model's generalization capabilities, i.e., how well it adapts to various network settings without significant re-training. A generalist model is built to handle such diversity, designed to perform robustly in different contexts. Achieving this broad applicability often relies on techniques like transfer learning or domain adaptation \cite{3GPP38.843, lin2023overview}. Transfer learning in optimization leverages insights or data from a previously solved problem (referred to as the \emph{source}) to accelerate the solution of a new, yet related problem (known as the \emph{target}). This approach is particularly beneficial when generating the initial dataset for the BO posterior is expensive or labor-intensive, such as when it requires experimental measurements.
In this section, we discuss the generalization capabilities of the HD-BO framework across different scenarios (i.e., UE distributions) within the context of \emph{transfer learning}.

{\begin{figure*}[!t]
    \centering
    \subfloat[Convergence of successful scenario-specific transfer learning]{
        \includegraphics[width=\figwidth]{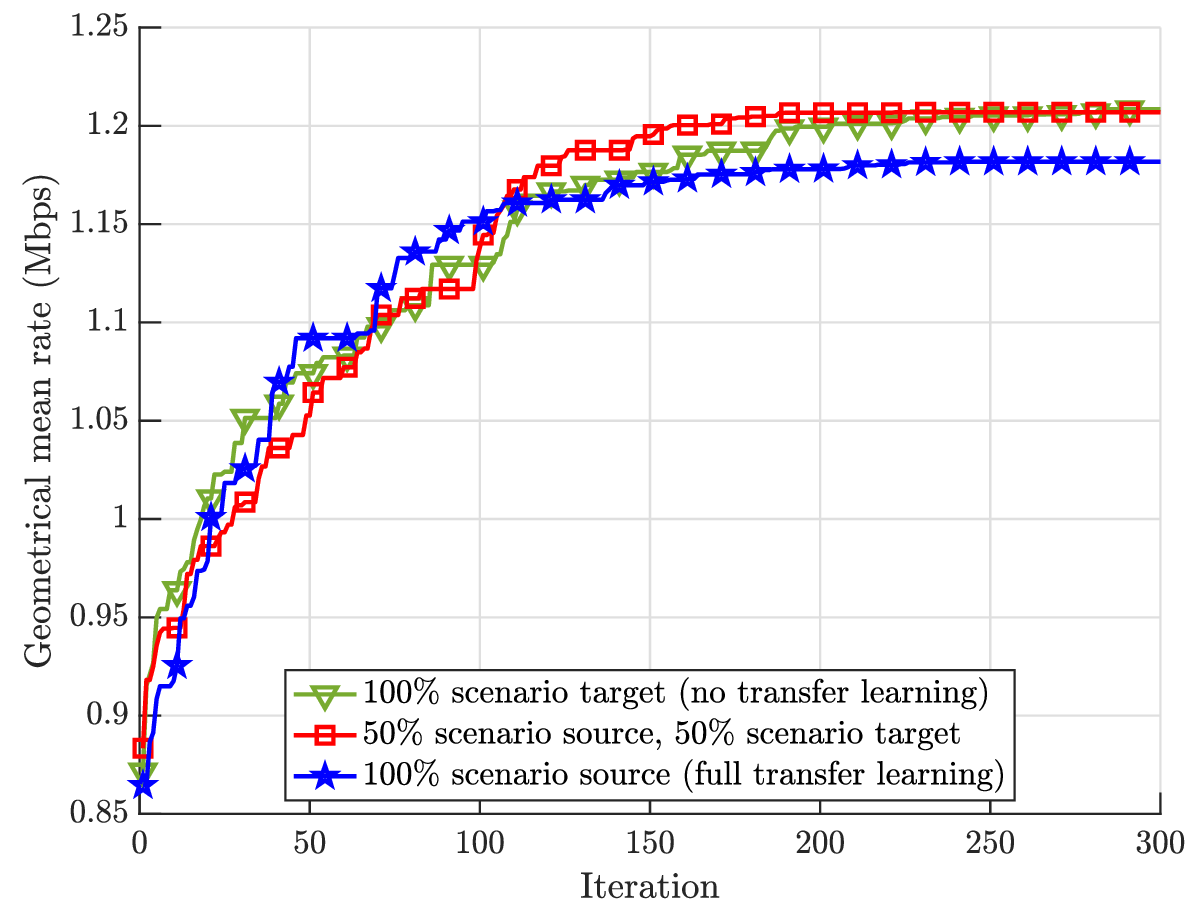}
        \label{fig:LT_conv_50m_150m}
    }
    \hspace{3mm}
    \subfloat[Convergence of unsuccessful scenario-specific transfer learning]{
        \includegraphics[width=\figwidth]{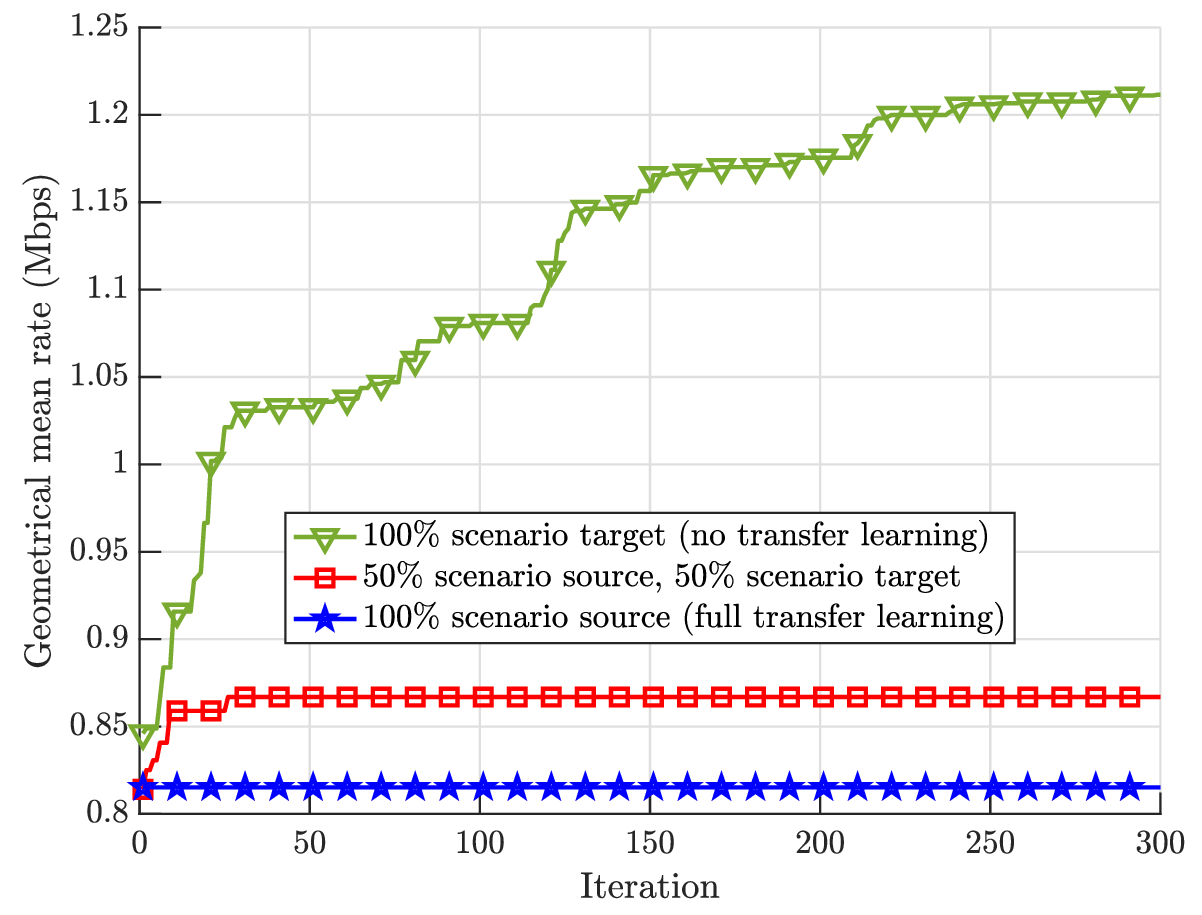}
        \label{fig:LT_conv_UAVs_NoUAVs}
    }
    \hspace{3mm}
    \subfloat[Performance of successful scenario-specific transfer learning]{
        \includegraphics[width=\figwidth]{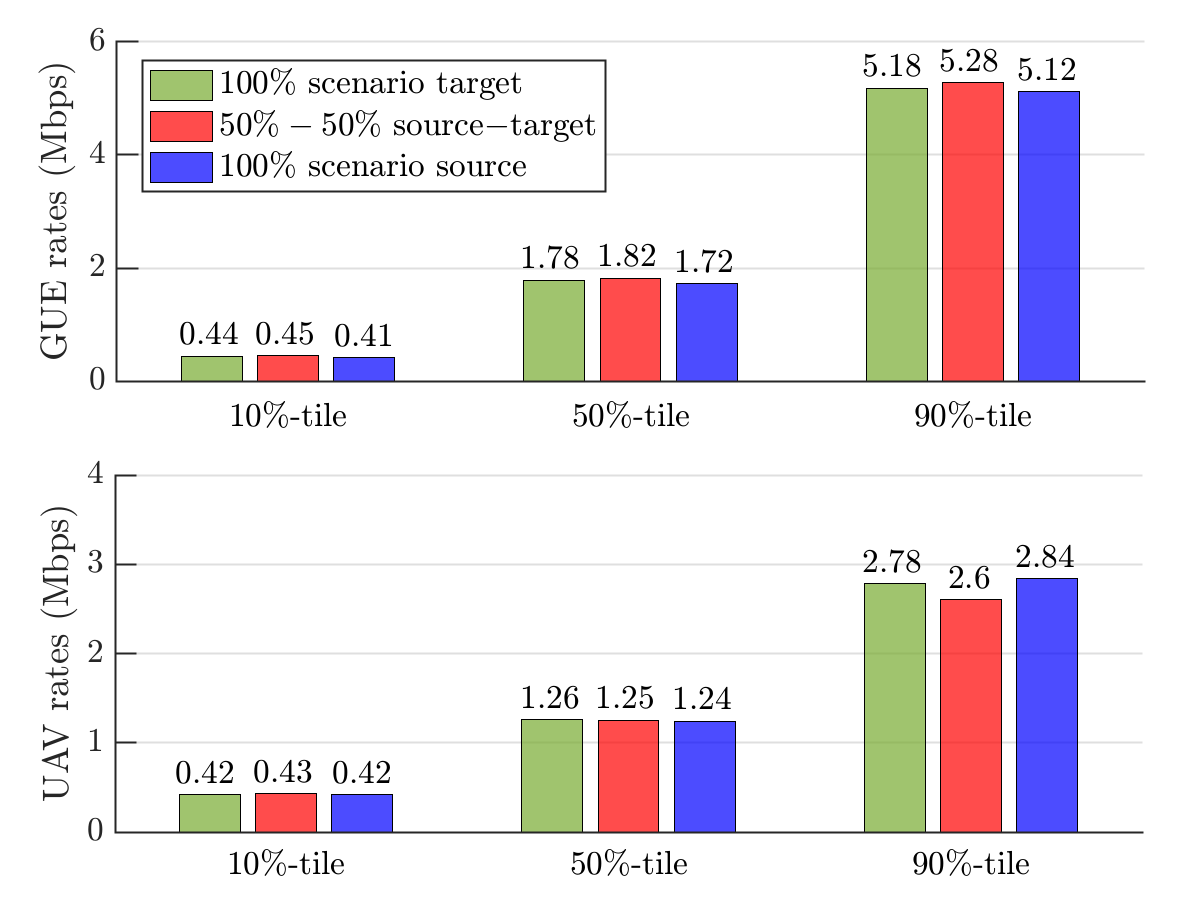}
        \label{fig:LT_compare_50m_150m_bar}
    }
    \hspace{3mm}
    \subfloat[Performance of unsuccessful scenario-specific transfer learning]{
        \includegraphics[width=\figwidth]{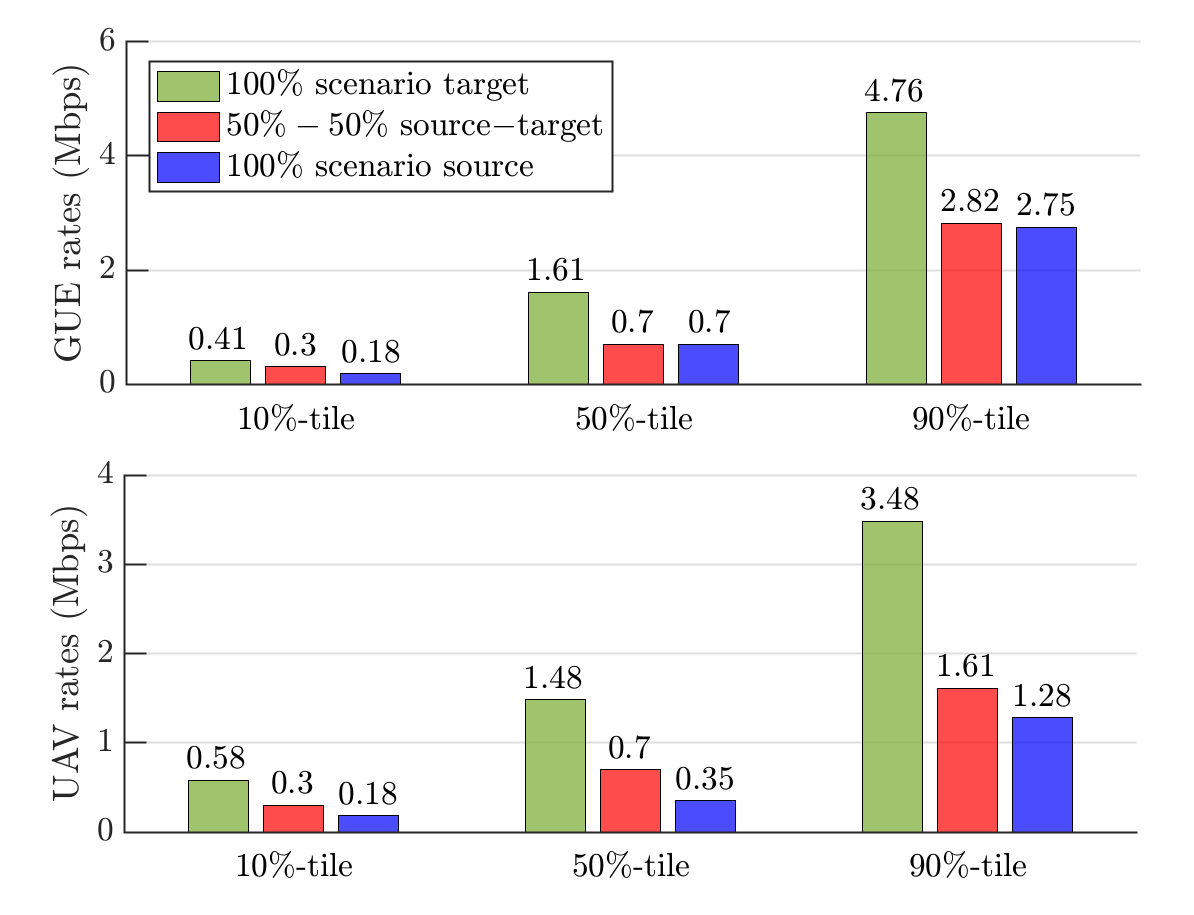}
        \label{fig:LT_compare_UAVs_NoUAVs}
    }
    \caption{Convergence (top) and performance (bottom) of scenario-specific transfer learning in two scenarios. (a,\,c) Source: UAVs corridors at 150\,m height. Target: UAVs corridors at 50\,m height. 
    (b,\,d) Source: No UAVs. Target: UAVs corridors at 150\,m height.}
    \label{fig:LT_compare_all}
\end{figure*}}



\subsection{Successful Scenario-specific Transfer Learning}

In \emph{scenario-specific} transfer learning, our goal is to use optimization data collected from applying the HD-BO framework to a specific aerial corridor deployment to optimize a new scenario where the altitude of the aerial corridors has changed. The scenario source is based on GUEs and UAVs along corridors at an altitude of 150\,m, whereas the scenario target changes the UAV corridor height to 50\,m.%
\footnote{
We found that \emph{configuration-specific} transfer learning is also possible: using data collected from a certain scenario to optimize a new one where a particular antenna parameter has changed. The details are omitted due to space constraints.
}

\subsubsection*{Convergence of scenario-specific transfer learning}

Figure~\ref{fig:LT_conv_50m_150m} illustrates the convergence of scenario-specific transfer learning using HD-BO, showing the best observed objective at each iteration $n$. Three evaluations were conducted, varying the initial dataset's dependency on the scenario target: 100\% scenario target (no transfer learning), 50\%, and 0\% (full transfer learning without any prior knowledge of the target optimization). Notably, with a 50\% reliance on the target's initial dataset, convergence occurs within a similar number of iterations, indicating that resources can be conserved when data sampling for the scenario target is costly. Even without prior knowledge of the target (initial dataset solely from the scenario source), performance declines by only a marginal 2\%.

\subsubsection*{Performance of scenario-specific transfer learning}

Figure~\ref{fig:LT_compare_50m_150m_bar} compares the achieved rates 
for both GUEs and UAVs. Comparable performance levels are observed across all three variations of the initial dataset for both GUEs and UAVs, demonstrating the effectiveness of transfer learning in accommodating UAVs in new corridor deployments while preserving performance for GUEs.


\subsection{Unsuccessful Scenario-specific Transfer Learning}

Here, we discuss an unsuccessful example where transfer learning proves ineffective for a scenario-specific category. Transfer learning is attempted for a scenario target involving GUEs and UAVs along corridors at a 150\,m altitude, but the scenario source includes only GUEs without any UAVs.


\subsubsection*{Convergence of scenario-specific transfer learning}

Fig.~\ref{fig:LT_conv_UAVs_NoUAVs} illustrates the convergence of scenario-specific transfer learning using the HD-BO algorithm, showing the best observed objective at each iteration $n$. This figure reveals that the HD-BO from the scenario source did not incorporate learning about UAVs. For instance, in cases where the initial dataset is 100\% from the scenario source, this limitation is apparent in the convergence plot, which displays constant performance throughout the optimization process. This occurs because TuRBO, the HD-BO method used, establishes trust regions based on presumed solution locations. Without UAV performance data, it lacks insight into the significance of antenna up-tilts and defaults to focusing only on down-tilts. Consequently, this approach fails to yield improvements in UAV corridor scenarios. In the 50\%-50\% case, although TuRBO recognizes the potential impact of up-tilts on UAVs, it struggles with accuracy due to the noisy data where up-tilting BSs decreased the KPIs in the other half of the dataset (i.e., the initial data containing only GUEs). As a result, TuRBO cannot determine the optimal trust regions for up-tilts and gets trapped in a low-performing local optimum.

\subsubsection*{Performance of scenario-specific transfer learning}

Fig.~\ref{fig:LT_compare_UAVs_NoUAVs} presents a comparison of the achieved rates 
for both GUEs and UAVs. This figure highlights the shortcomings of the initial-dataset variations tested (50\%-50\% and 100\% scenario source), as in both cases the rate performance of GUEs and UAVs fails to reach the performance achieved when the optimization is conducted with an initial dataset composed entirely of the scenario target (100\% scenario target).

\section{Capacity-coverage Tradeoff via Multi-\\objective High-dimensional Bayesian Optimization}
\label{sec:MORBO}


Simultaneous optimization of capacity and coverage is a common multi-objective problem in the design of cellular networks. The contrasting nature of these two objectives is exacerbated when designing networks for UAV connectivity. Reliable coverage for UAVs requires directing energy upwards by up-tilting some antenna sectors, which reduces performance for GUEs which benefit from down-tilted sectors. Multi-objective problems are typically addressed by merging two objectives through a linear combination, e.g., with the mixing ratio $\lambda$ discussed in Section \ref{sec:BO} \cite{dreifuerst2021optimizing}. However, this method does not clearly reveal the trade-offs between objectives. Therefore, a multi-objective framework that defines the Pareto front between two contrasting  objectives (i.e., optimizing for capacity and coverage) is desirable. In this section, we discuss the generalization of the HD-BO framework to multi-objective problems and apply it to optimize the BS antenna tilts to achieve a trade-off between UAV coverage and GUE data rates. Unlike population-based methods such as NSGA-II or MOEA/D~\cite{pan2023joint}, which typically require a large number of evaluations to approximate the Pareto front, MORBO offers a sample-efficient alternative tailored for expensive black-box functions. It enables the discovery of Pareto-optimal solutions with significantly fewer iterations, making it well-suited for the formulated problem.


\subsection{Introduction to Multi-objective HD-BO} 

In multi-objective optimization, the goal is to maximize a vector-valued objective function $\mathbf{f}(\mathbf{x}) = \left[ f^{(1)}(\mathbf{x}), \ldots, f^{(M)}(\mathbf{x}) \right] \in \mathbb{R}^M$, where 
$M \geq 2$, $\mathbf{x} \in \bm{\mathcal{X}} \subset \mathbb{R}^d$, and $\bm{\mathcal{X}}$ is a compact set. Typically, no single solution $\mathbf{x}^{*}$ can maximize all $M$ objectives at once; therefore, objective vectors are evaluated based on Pareto domination \cite{daulton2022multi}. An objective vector $\mathbf{f}(\mathbf{x})$ \textit{Pareto-dominates} $\mathbf{f}(\mathbf{x'})$, denoted as $\mathbf{f}(\mathbf{x}) \succ \mathbf{f}(\mathbf{x'})$, if $f^{(m)}(\mathbf{x}) \geq f^{(m)}(\mathbf{x'})$ for all $m = 1, \ldots, M$ and there exists at least one $m$, such that $f^{(m)}(\mathbf{x}) > f^{(m)}(\mathbf{x'})$. The \textit{Pareto front} (PF) is the set of optimal trade-offs $P(\bm{X})$ over a set of configurations $\bm{X} \subseteq \bm{\mathcal{X}}$ \cite{daulton2022multi}:
\begin{equation}
P(\bm{X}) = \{\mathbf{f}(\mathbf{x}) : \mathbf{x} \in \bm{X}, \nexists \, \mathbf{x'} \in \bm{X} \text{ s.t. } \mathbf{f}(\mathbf{x'}) \succ \mathbf{f}(\mathbf{x})\}.
\end{equation}
The goal of a multi-objective optimization algorithm is to identify an approximate PF $P(\bm{X}_n)$ of the true PF $P(\bm{X})$ within a specified budget of $ \lvert \bm{X}_n \rvert = n$ function evaluations. 

\subsubsection*{Hypervolume improvement}

The quality of a PF is evaluated by the hypervolume (HV) indicator. The \textit{hypervolume indicator}, HV $(P(\bm{X}) | \bm{r})$ is the $M$-dimensional Lebesgue measure $\lambda_{M}$ of the region
dominated by $P(\bm{X})$ and bounded from below by a reference point $\bm{r} \in \mathbb{R}^M$, the latter typically provided based on domain knowledge \cite{yang2019multi}. The \textit{hypervolume improvement} (HVI) from a set of points is the increase in HV when adding these points to the previously selected points. Maximizing the hypervolume improvement with a given reference point has been demonstrated to generate PFs of high quality and diversity \cite{emmerich2006single}. 

\subsubsection*{Multi-objective Bayesian Optimization over High-dimensional Search Spaces (MORBO)}

MORBO \cite{daulton2022multi} is a collaborative multi-TR approach for high-dimensional multi-objective BO that takes a different approach compared to TuRBO. Instead of using multiple independent trust regions, MORBO shares observations across them. This means that each TR is provided with all available information regarding the objectives relevant for local optimization within the TR. Thus, MORBO coordinates the selection of TR center points to encourage identifying Pareto fronts with comprehensive coverage of the search space, selecting new candidate designs by collaboratively optimizing a shared global utility function, and utilizing local models to decrease computational complexity and enhance scalability, particularly in scenarios involving large amounts of data. 
Thompson sampling is employed to draw $q$ posterior samples from the GP and then optimize the HVI under each realization. This strategy can be seen as a single-sample approximation of the expected HVI (EHVI), a widely used acquisition function that integrates HVI over the GP posterior \cite{daulton2021parallel}. 

\subsubsection*{Trust region placement}

In constrained single-objective optimization, prior research typically places the local TR at the best observed point, usually the best feasible solution. However, in the multi-objective context, there is often no single best solution. MORBO selects the center of the TR based on the feasible point on the PF that offers the maximum hypervolume contribution (HVC). Using a reference point, the HVC of a point on the PF measures the decrease in hypervolume (HV) that would result if that point were removed; essentially, the HVC represents the unique contribution of that point to the PF. Centering a TR at the point with the maximum HVC collected by that TR helps enhance coverage across the PF. This way, points in densely populated areas of the PF tend to have lower contributions, thus promoting exploration of less crowded regions. 

\subsubsection*{Computational complexity}

As MORBO shares observations across TRs and employs local models, this approach
significantly reduces the computational cost of the algorithm since exact GP
fitting scales cubically with the number of data points. Using local models results in speedups of $\mathcal{O}\left(n_{TR}^2 / \eta^3\right)$, where $n_{TR}$ represents the sampled points within a TR, and $\eta$ is the average number of TR modeling spaces a data point resides in. This may translates into speedups of two orders of magnitude relative to global modeling, as shown in \cite[Appendix F.2.]{daulton2022multi}.


\subsection{Trading-off UAV Coverage with GUE Capacity via MORBO}

We now present the results obtained when applying the MORBO framework to the antenna tilt optimization problem (\ref{eqn:Opt_problem}) with the following two objectives:
\begin{itemize}[leftmargin=*]
\item
The UAV probability of coverage, $P_{\text{cov}}^{\text{UAV}}$, defined as the fraction of UAVs with SINR above a threshold $\tau$, i.e.:
\begin{equation}
P_{\text{cov}}^{\text{UAV}} = \frac{\sum_{k \in \ncalU} \mathbbm{1}_{\{ \sinr_{\textrm{dB},k} \geq \tau \}}}{\|\ncalU\|}
\end{equation}
where $\mathbbm{1}_{(\cdot)}$ denotes the indicator function and $\|\ncalU\|$ is the cardinality of the set of UAV UEs, $\ncalU$.
\item 
The GUE sum-log-rates, obtained from (\ref{eqn:Opt_problem}) when $\lambda = 0$.
\end{itemize}
The two objectives contrast in nature as enhancing the UAV coverage probability requires some BSs to be up-tilted, thus moving GUEs which were associated with those BSs to other BSs, whose increased load causes a lower per-GUE rate.

We implement MORBO using BoTorch \cite{balandat2020botorch} with five trust regions. Following \cite{eriksson2019scalable}, we set the hyperparameters of the trust regions to: $L_{\text{init}} = 0.8, L_{\text{min}} = 0.01, L_{\text{max}} = 1.6$. For each trust region, we use an independent GP with a constant mean function and a Matérn-5/2 kernel. The GP hyperparameters are fit by maximizing the marginal log-likelihood.

\begin{figure}[!t]
    \centering
    \includegraphics[width=\figwidth]{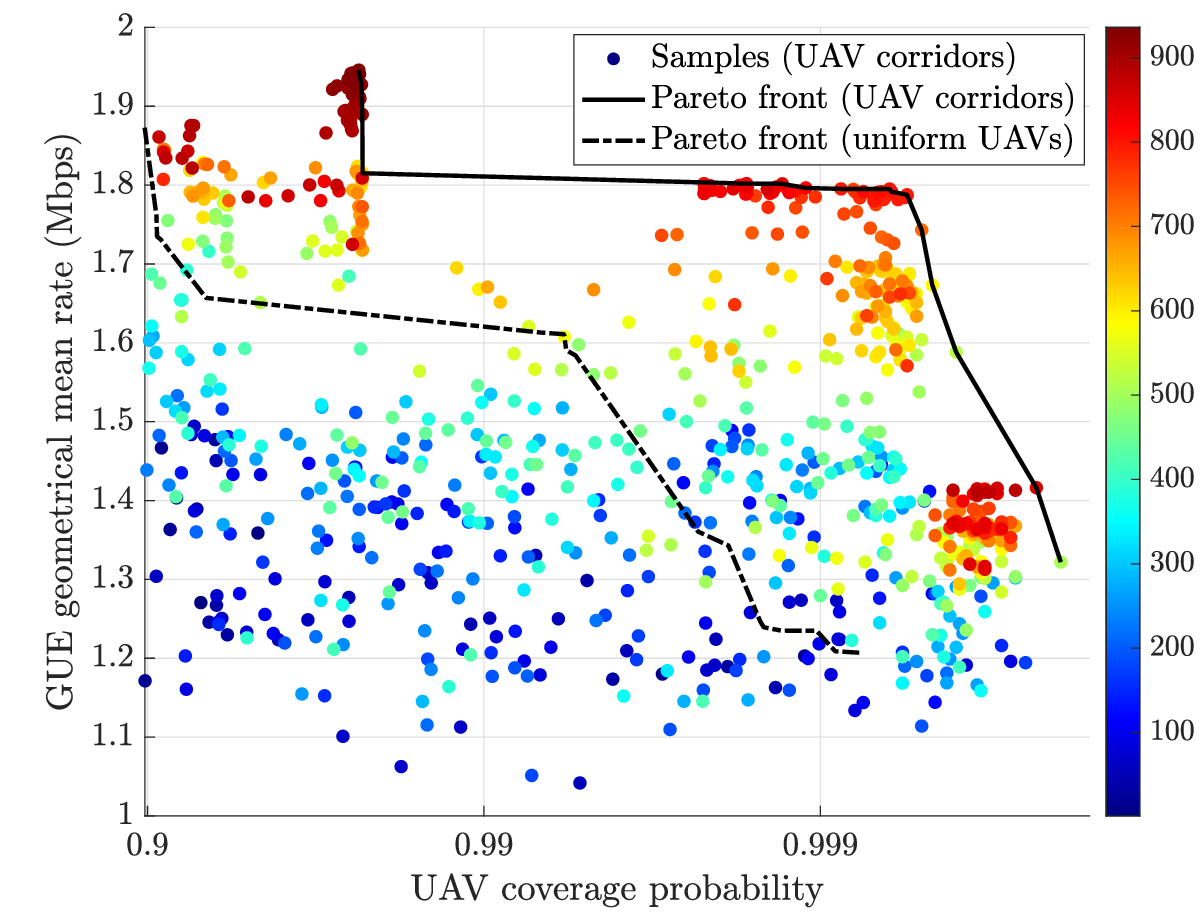}
    \caption{Search space with MORBO, with the color bar indicating the iteration number. Black lines depict the Pareto fronts obtained under UAV corridors (solid) and UAVs uniformly distributed (dash-dot).}
    \label{fig:MORBO_PcUAVs_RatesGUEs_all}
\end{figure}

Fig.~\ref{fig:MORBO_PcUAVs_RatesGUEs_all} shows results obtained when the UAV coverage threshold is set to $\tau = -5$\,dB in scenarios with UAV corridors, with the color bar indicating the iteration number. The solid and dash-dot black line depict the Pareto fronts obtained in scenarios with UAV corridors and uniformly distributed UAVs, respectively. 
This multi-objective optimization study confirms our previous findings regarding the advantages of UAV corridors in terms of ground-plus-air connectivity optimization. 
For instance, when UAVs are uniformly distributed---and thus UAV coverage is pursued across the entire sky---, increasing the reliability from two (0.99) to three (0.999) `nines' comes at a cost of decreasing the GUE geometrical mean rate from 1.6 to 1.2\,Mbps. 
When UAVs are confined to corridors, increasing the coverage probability to three nines is possible while ensuring a GUE geometrical mean rate of 1.8\,Mbps. Similarly, for a given GUE geometrical mean rate of 1.3\,Mbps, the UAV coverage reliability with uniform UAVs and UAV corridors is of 0.998 and 0.9998, respectively, corresponding to a one order of magnitude reduction in outage.

\section{Case Study: Real-world Cellular Network Topology and Site-specific Propagation Channel}
\label{sec:RD-HDBO}

In the previous sections, we studied the capability of HD-BO in optimizing cell antenna parameters employing a 3GPP statistical channel model for a conventional hexagonal cellular layout. In this section, we consider a site-specific scenario corresponding to a real-world cellular network \cite{benzaghta2025wcnc}. Our problem formulation is similar to (\ref{eqn:Opt_problem_joint}) in Section~\ref{sec:HD-BO}, with the scenario detailed as follows:

\subsubsection*{Real cellular network topology} 

\begin{figure}
\centering
\includegraphics[width=\figwidth]{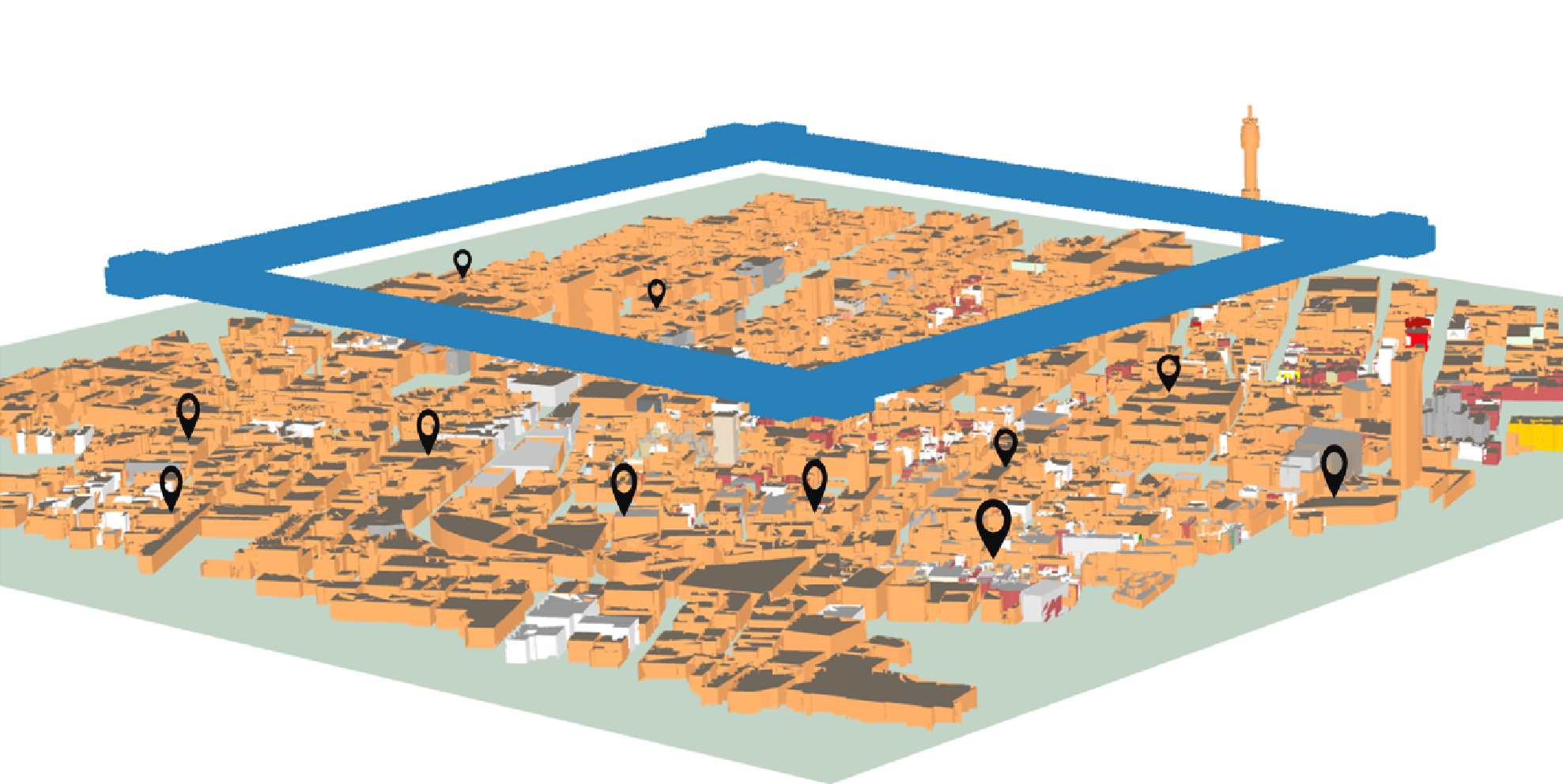}
\caption{Area of London considered, with some of the cell deployment sites indicated by black markers and 3D UAV corridors in blue.}
\label{fig:RD_illustration}
\vspace*{-0.2cm}
\end{figure}

For our real-world case study, we consider a production radio network owned by a leading commercial mobile operator in the UK. The portion of the network considered consists of 16 deployment sites varying in height from 22 to 56\,m and operating a carrier in the 2\,GHz band spectrum. Each site is equipped with three sector antennas, for a total of 48 cells. The geographical area selected for our study spans approximately 1400\,m by 1275\,m, and is located within London, between latitude $[51.5087, 51.5215]$ and longitude $[-0.1483, -0.1296]$. Within this area, GUEs are randomly located outdoor (i.e., not within buildings) at a height of 1.5\,m, with a density of 10 GUEs per cell on average \cite{3GPP36777}. We also consider four 3D aerial corridors within the area, each 900\,m in length, 40\,m wide, and positioned at heights between 140\,m to 160\,m. The UAVs-to-GUEs ratio is set to 50\% following the 3GPP Case~5 in \cite{3GPP36777}. Fig.~\ref{fig:RD_illustration} illustrates a 3D model of the selected area in London, identifying some of the cell sites locations, and showing a representation of the 3D aerial corridors. Optimizing the cell antenna parameters within this urban scenario is a non-trivial task, as different areas present different signal propagation patterns. This brings the challenge of tailoring the deployment to the location characteristics while accounting for interference and load balancing. The need to also provide reliable connectivity along UAV corridors further complicates the problem, requiring configurations that are not easily designed heuristically.

\subsubsection*{Site-specific propagation channel} 

A 3D representation of the neighborhood we selected is constructed from OpenStreetMap, including terrain and building information. BSs are positioned and configured as per the real cellular network topology. The large-scale channel gain $G_{b,k}$ (not including the antenna gain) between BS $b$ and UE $k$, is obtained using Sionna RT \cite{hoydis2023sionna}, a 3D ray-tracing tool widely used for analyzing site-specific radio wave propagation. Simulations are conducted at the operating carrier frequency of 2\,GHz. 
The material \texttt{itu_concrete} is used to model the permittivity and conductivity of all buildings, and the maximum number of reflections and diffractions are set to 5 and 1, respectively. The total large-scale channel gain is then obtained from the omnidirectional ray tracing channel gain by adding the antenna gain as per (\ref{eqn: AG}). 


\subsubsection*{Performance of HD-BO}

Table~\ref{tab:RD_performance} presents a comparison of the performance achieved by GUEs and UAVs when the cellular network is optimized for both GUEs and UAVs ($\lambda = 0.5$), by jointly tuning the antenna tilts and HPBW with TuRBO. The performance of the cellular network with its original configuration of cell tilts and HPBW is also reported as a baseline for comparison. 
The recommended configurations by TuRBO for uniform deployment include 29 up-tilted cells out of 48, while this number reduces to 22 up-tilted cells for corridors deployment. 
In the baseline scenario, where all cells are down-tilted within the range $[-12^{\circ}, -4^{\circ}]$, nearly half of the UAVs experience outages. This falls to 8\% for uniform deployments and 2\% for corridors deployment when using our optimization. TuRBO also enhances the UAV's geometric mean rate by 90\% for the uniform deployment and 96\% for corridors deployment. 
This UAV performance improvement correspond to only a slight GUE performance degradation: 6\% for uniform deployment and 3\% for corridors deployment, demonstrating the effectiveness of the optimization procedure for real cellular networks.

\begin{table}[h]
\centering
\caption{KPIs with a baseline configuration and with an optimized one for uniformly distributed UAVs (`Uni') and UAV corridors (`Corr'). }
\begin{tabular}{l c c c}
\hline
 & \textbf{Baseline} & \textbf{Uni} & \textbf{Corr} \\
\hline
Number of up-tilted cells & 0 & 29 & 22 \\
\hline
UAVs coverage probability ($\tau = -5$\,dB) & 0.505 & 0.923 & 0.985 \\
\hline
Geometric mean rate, GUEs + UAVs (Mbps) & 2.60 & 2.97 & 3.30 \\
\hline
Geometric mean rate, GUEs (Mbps) & 3.24 & 3.04 & 3.15 \\
\hline
Geometric mean rate, UAVs (Mbps) & 0.48 & 0.91 & 0.94 \\
\hline
\end{tabular}
\label{tab:RD_performance}
\end{table}

\section{Conclusion and Future Directions}
\label{sec:conclusion}


\subsection{Summary of Results}

In this paper, we tackled the challenge of designing cellular networks for UAV corridors using a novel data-driven approach with HD-BO. Our key takeaways are summarized as follows:

\begin{itemize}[leftmargin=*]

\item 
\emph{Optimal design for UAV corridors:}
An optimal combination of antenna tilts and vHPBW significantly enhances performance along UAV corridors. This optimized setup achieves gains exceeding 20\,dB in median SINR over an all-downtilt, fixed-HPBW 3GPP baseline, with only a marginal reduction in ground performance compared to scenarios without UAVs. Notably, fewer cells are required to cover UAVs along corridors, resulting in a 40\% increase in mean rate compared to uniformly distributed UAVs at a height 150\,m.

\item
\emph{Efficiency and convergence of HD-BO:}
Inspired by traditional BO, HD-BO constructs data-driven models that efficiently narrow down the space of candidate solutions. TuRBO converges within 300 iterations across all tested scenarios, showing remarkable superiority compared to other benchmarks.

\item
\emph{Transfer learning:}
HD-BO enables transfer learning across scenarios, utilizing existing data to conserve resources and maintain performance without significant degradation. When informed about UAV presence, HD-BO adapts effectively between scenarios. Conversely, without prior UAV exposure, HD-BO may fail to identify optimal regions for antenna adjustments, leading to suboptimal outcomes.

\item
\emph{Coverage-capacity trade-offs:}
Multi-objective HD-BO effectively balances trade-offs between GUE data rates and UAV coverage reliability. UAV corridors achieve far superior trade-offs compared to uniform UAV distribution, where achieving high UAV reliability reduces GUE rates more significantly.

\item
\emph{Performance in a real-world deployment:}
Our case study in a production cellular network confirms the findings from 3GPP statistical models. HD-BO identifies optimal, non-obvious antenna configurations that can achieve more than double UAV rates while incurring a negligible GUE performance reduction.

\end{itemize}


\subsection{Limitations and Future Work}

This paper primarily focused on optimizing antenna tilts and HPBW. However, with the evolution towards 5G and beyond, it becomes crucial to consider synchronization signal blocks (SSBs) beams \cite{LopPioGer2024}. 
Future work could explore the data-driven optimization of beam codebooks across multiple cells. 
Furthermore, future extensions could incorporate user mobility, allowing the HD-BO framework to optimize handover (HO) parameters, such as the time-of-trigger and A3 offset, to improve network performance by reducing HO failures \cite{benzaghta2025pimrc}. 

\bibliographystyle{IEEEtran}
\bibliography{journalAbbreviations, main}

\begin{thebibliography}{10}
\providecommand{\url}[1]{#1}
\csname url@samestyle\endcsname
\providecommand{\newblock}{\relax}
\providecommand{\bibinfo}[2]{#2}
\providecommand{\BIBentrySTDinterwordspacing}{\spaceskip=0pt\relax}
\providecommand{\BIBentryALTinterwordstretchfactor}{4}
\providecommand{\BIBentryALTinterwordspacing}{\spaceskip=\fontdimen2\font plus
\BIBentryALTinterwordstretchfactor\fontdimen3\font minus
  \fontdimen4\font\relax}
\providecommand{\BIBforeignlanguage}[2]{{%
\expandafter\ifx\csname l@#1\endcsname\relax
\typeout{** WARNING: IEEEtran.bst: No hyphenation pattern has been}%
\typeout{** loaded for the language `#1'. Using the pattern for}%
\typeout{** the default language instead.}%
\else
\language=\csname l@#1\endcsname
\fi
#2}}
\providecommand{\BIBdecl}{\relax}
\BIBdecl

\bibitem{benzaghta2023designing}
M.~Benzaghta, G.~Geraci, D.~L{\'o}pez-P{\'e}rez, and A.~Valcarce, ``Designing
  cellular networks for {UAV} corridors via {B}ayesian optimization,'' in
  \emph{Proc. IEEE Globecom}, 2023, pp. 4552--4557.

\bibitem{Wu2021}
Q.~Wu, J.~Xu, Y.~Zeng, D.~W.~K. Ng, N.~Al-Dhahir, R.~Schober, and A.~L.
  Swindlehurst, ``A comprehensive overview on {5G}-and-beyond networks with
  {UAVs}: {F}rom communications to sensing and intelligence,'' \emph{{IEEE} J.
  Sel. Areas Commun.}, vol.~39, no.~10, pp. 2912--2945, 2021.

\bibitem{GerGarAza2022}
G.~Geraci, A.~Garcia-Rodriguez, M.~M. Azari, A.~Lozano, M.~Mezzavilla,
  S.~Chatzinotas, Y.~Chen, S.~Rangan, and M.~Di~Renzo, ``What will the future
  of {UAV} cellular communications be? {A} flight from {5G} to {6G},''
  \emph{{IEEE} Commun. Surveys Tuts.}, vol.~24, no.~3, pp. 1304--1335, 2022.

\bibitem{FotQiaDin2019}
A.~{Fotouhi}, H.~{Qiang}, M.~{Ding}, M.~{Hassan}, L.~{Galati-Giordano},
  A.~{Garcia-Rodriguez}, and J.~{Yuan}, ``Survey on {UAV} cellular
  communications: {P}ractical aspects, standardization advancements,
  regulation, and security challenges,'' \emph{{IEEE} Commun. Surveys Tuts.},
  vol.~21, no.~4, pp. 3417--3442, 2019.

\bibitem{zeng2020uav}
Y.~Zeng, I.~Guvenc, R.~Zhang, G.~Geraci, and D.~W. Matolak, \emph{{UAV}
  Communications for {5G} and Beyond}.\hskip 1em plus 0.5em minus 0.4em\relax
  John Wiley \& Sons, 2020.

\bibitem{GiuNikGer2024}
A.~Giuliani, R.~Nikbakht, G.~Geraci, S.~Kang, A.~Lozano, and S.~Rangan,
  ``Spatially consistent air-to-ground channel modeling via generative neural
  networks,'' \emph{{IEEE} Commun. Lett.}, vol.~13, no.~4, pp. 1158--1162,
  2024.

\bibitem{geraci2018understanding}
G.~Geraci, A.~Garcia-Rodriguez, L.~Galati-Giordano, D.~L{\'o}pez-P{\'e}rez, and
  E.~Bj{\"o}rnson, ``Understanding {UAV} cellular communications: From existing
  networks to massive {MIMO},'' \emph{IEEE Access}, vol.~6, pp.
  67\,853--67\,865, 2018.

\bibitem{ZenLyuZha2019}
Y.~Zeng, J.~Lyu, and R.~Zhang, ``Cellular-connected {UAV}: {P}otentials,
  challenges and promising technologies,'' \emph{IEEE Wireless Commun.},
  vol.~26, no.~1, pp. 120--127, 2019.

\bibitem{3GPP36777}
{3GPP Technical Report 36.777}, ``{S}tudy on enhanced {LTE} support for aerial
  vehicles ({R}elease 15),'' Dec. 2017.

\bibitem{NguAmoWig2018}
{H.~C.~Nguyen}, R.~Amorim, J.~Wigard, {I.~Z.~Kov\'{a}cs}, {T.~B.~S{\o}rensen},
  and {P.~Mogensen}, ``How to ensure reliable connectivity for aerial vehicles
  over cellular networks,'' \emph{IEEE Access}, vol.~6, pp. 12\,304--12\,317,
  2018.

\bibitem{pan2023resource}
H.~Pan, Y.~Liu, G.~Sun, P.~Wang, and C.~Yuen, ``Resource scheduling for
  {UAVs}-aided {D2D} networks: {A} multi-objective optimization approach,''
  \emph{IEEE Transactions on Wireless Communications}, 2023.

\bibitem{KanMezLoz2021}
S.~Kang, M.~Mezzavilla, A.~Lozano, G.~Geraci, W.~Xia, S.~Rangan, V.~Semkin, and
  G.~Loianno, ``Millimeter-wave {UAV} coverage in urban environments,'' in
  \emph{Proc. IEEE Globecom}, 2021.

\bibitem{XiaRanMez2020a}
W.~Xia, S.~Rangan, M.~Mezzavilla, A.~Lozano, G.~Geraci, V.~Semkin, and
  G.~Loianno, ``Generative neural network channel modeling for millimeter-wave
  {UAV} communication,'' \emph{IEEE Trans. Wireless Commun.}, vol.~21, no.~11,
  pp. 9417--9431, 2022.

\bibitem{kim2022non}
S.~Kim, M.~Kim, J.~Y. Ryu, J.~Lee, and T.~Q. Quek, ``{Non-terrestrial networks
  for UAVs: Base station service provisioning schemes with antenna tilt},''
  \emph{IEEE Access}, vol.~10, pp. 41\,537--41\,550, 2022.

\bibitem{mozaffari2021toward}
M.~Mozaffari, X.~Lin, and S.~Hayes, ``{Toward {6G} with connected sky: {UAVs}
  and beyond},'' \emph{IEEE Commun. Mag.}, vol.~59, no.~12, pp. 74--80, 2021.

\bibitem{d2020analysis}
C.~D’Andrea, A.~Garcia-Rodriguez, G.~Geraci, L.~G. Giordano, and S.~Buzzi,
  ``{Analysis of {UAV} communications in cell-free massive {MIMO} systems},''
  \emph{IEEE Open J. of the Commun. Society}, vol.~1, pp. 133--147, 2020.

\bibitem{garcia2019essential}
A.~Garcia-Rodriguez, G.~Geraci, D.~L{\'o}pez-P{\'e}rez, L.~G. Giordano,
  M.~Ding, and E.~Bjornson, ``The essential guide to realizing {5G}-connected
  {UAVs} with massive {MIMO},'' \emph{IEEE Commun. Mag.}, vol.~57, no.~12, pp.
  84--90, 2019.

\bibitem{benzaghta2022uav}
M.~Benzaghta, G.~Geraci, R.~Nikbakht, and D.~L{\'o}pez-P{\'e}rez, ``{UAV}
  communications in integrated terrestrial and non-terrestrial networks,'' in
  \emph{Proc. IEEE Globecom}, 2022, pp. 3706--3711.

\bibitem{geraci2022integrating}
G.~Geraci, D.~L{\'o}pez-P{\'e}rez, M.~Benzaghta, and S.~Chatzinotas,
  ``Integrating terrestrial and non-terrestrial networks: {3D} opportunities
  and challenges,'' \emph{IEEE Commun. Mag.}, 2023.

\bibitem{esrafilian2020three}
O.~Esrafilian, R.~Gangula, and D.~Gesbert, ``Three-dimensional-map-based
  trajectory design in {UAV}-aided wireless localization systems,'' \emph{IEEE
  Internet of Things Journal}, vol.~8, no.~12, pp. 9894--9904, 2020.

\bibitem{challita2018deep}
U.~Challita, W.~Saad, and C.~Bettstetter, ``Deep reinforcement learning for
  interference-aware path planning of cellular-connected {UAVs},'' in
  \emph{Proc. IEEE ICC}, 2018, pp. 1--7.

\bibitem{de2019cellular}
S.~De~Bast, E.~Vinogradov, and S.~Pollin, ``Cellular coverage-aware path
  planning for {UAVs},'' in \emph{Proc. IEEE SPAWC}, 2019, pp. 1--5.

\bibitem{matracia2023uav}
M.~Matracia, M.~A. Kishk, and M.-S. Alouini, ``{UAV}-aided post-disaster
  cellular networks: {A} novel stochastic geometry approach,'' \emph{IEEE
  Trans. on Vehicular Technology}, vol.~72, no.~7, pp. 9406--9418, 2023.

\bibitem{galkin2019stochastic}
B.~Galkin, J.~Kibi{\l}da, and L.~A. DaSilva, ``A stochastic model for {UAV}
  networks positioned above demand hotspots in urban environments,'' \emph{IEEE
  Trans. on Vehicular Technology}, vol.~68, no.~7, pp. 6985--6996, 2019.

\bibitem{cherif20213d}
N.~Cherif, W.~Jaafar, H.~Yanikomeroglu, and A.~Yongacoglu, ``{3D} aerial
  highway: {T}he key enabler of the retail industry transformation,''
  \emph{IEEE Commun. Mag.}, vol.~59, no.~9, pp. 65--71, 2021.

\bibitem{bhuyan2021secure}
A.~Bhuyan, {\.I}.~G{\"u}ven{\c{c}}, H.~Dai, M.~L. Sichitiu, S.~Singh,
  A.~Rahmati, and S.~J. Maeng, ``Secure {5G} network for a nationwide drone
  corridor,'' in \emph{Proc. IEEE Aerospace Conference}, 2021, pp. 1--10.

\bibitem{bulut2018trajectory}
E.~Bulut and I.~Guvenc, ``Trajectory optimization for cellular-connected {UAVs}
  with disconnectivity constraint,'' in \emph{Proc. IEEE ICC Workshops}, 2018,
  pp. 1--6.

\bibitem{bayerlein2021multi}
H.~Bayerlein, M.~Theile, M.~Caccamo, and D.~Gesbert, ``Multi-{UAV} path
  planning for wireless data harvesting with deep reinforcement learning,''
  \emph{IEEE Open J. of the Commun. Society}, vol.~2, pp. 1171--1187, 2021.

\bibitem{bernabe2022optimization}
M.~Bernab{\`e}, D.~Lopez-Perez, D.~Gesbert, and H.~Bao, ``On the optimization
  of cellular networks for {UAV} aerial corridor support,'' in \emph{Proc. IEEE
  Globecom}, 2022, pp. 2969--2974.

\bibitem{bernabe2023novel}
M.~Bernab{\`e}, D.~L{\'o}pez-P{\'e}rez, N.~Piovesan, G.~Geraci, and D.~Gesbert,
  ``A novel metric for {mMIMO} base station association for aerial highway
  systems,'' in \emph{Proc. IEEE ICC Workshops}, 2023, pp. 1063--1068.

\bibitem{maeng2023base}
S.~J. Maeng, M.~M.~U. Chowdhury, {\.I}.~G{\"u}ven{\c{c}}, A.~Bhuyan, and
  H.~Dai, ``Base station antenna uptilt optimization for cellular-connected
  drone corridors,'' \emph{IEEE Trans. on Aerospace and Electronic Systems},
  2023.

\bibitem{chowdhury2021ensuring}
M.~M.~U. Chowdhury, I.~Guvenc, W.~Saad, and A.~Bhuyan, ``Ensuring reliable
  connectivity to cellular-connected {UAVs} with up-tilted antennas and
  interference coordination,'' \emph{arXiv:2108.05090}, 2021.

\bibitem{karimi2023optimizing}
S.~Karimi-Bidhendi, G.~Geraci, and H.~Jafarkhani, ``Optimizing cellular
  networks for {UAV} corridors via quantization theory,'' \emph{IEEE Trans.
  Wireless Commun.}, 2024.

\bibitem{karimi2023analysis}
S.~Karimi-Bidhendi, G.~Geraci, and H.~Jafarkhani, ``Analysis of {UAV} corridors
  in cellular networks,'' in \emph{Proc. IEEE ICC}, 2023, pp. 1--6.

\bibitem{shahriari2015taking}
B.~Shahriari, K.~Swersky, Z.~Wang, R.~P. Adams, and N.~De~Freitas, ``Taking the
  human out of the loop: {A} review of {Bayesian} optimization,''
  \emph{Proceedings of the IEEE}, vol. 104, no.~1, pp. 148--175, 2015.

\bibitem{dreifuerst2021optimizing}
R.~M. Dreifuerst, S.~Daulton, Y.~Qian, P.~Varkey, M.~Balandat, S.~Kasturia,
  A.~Tomar, A.~Yazdan, V.~Ponnampalam, and R.~W. Heath, ``Optimizing coverage
  and capacity in cellular networks using machine learning,'' in \emph{Proc.
  IEEE ICASSP}, 2021, pp. 8138--8142.

\bibitem{eller2024differentiable}
L.~Eller, P.~Svoboda, and M.~Rupp, ``A differentiable throughput model for
  load-aware cellular network optimization through gradient descent,''
  \emph{IEEE Access}, 2024.

\bibitem{zhang2023bayesian}
Y.~Zhang, O.~Simeone, S.~T. Jose, L.~Maggi, and A.~Valcarce, ``Bayesian and
  multi-armed contextual meta-optimization for efficient wireless radio
  resource management,'' \emph{IEEE Trans. on Cognitive Communications and
  Networking}, 2023.

\bibitem{maggi2021bayesian}
L.~Maggi, A.~Valcarce, and J.~Hoydis, ``Bayesian optimization for radio
  resource management: {O}pen loop power control,'' \emph{IEEE J. Sel. Areas
  Commun.}, vol.~39, no.~7, pp. 1858--1871, 2021.

\bibitem{tambovskiy2022cell}
S.~S. Tambovskiy, G.~Fodor, and H.~Tullberg, ``Cell-free data power control via
  scalable multi-objective {B}ayesian optimisation,'' in \emph{Proc. IEEE
  PIMRC}, 2022, pp. 1--6.

\bibitem{maggi2023energy}
L.~Maggi, C.~Mihailescu, Q.~Cao, A.~Tetich, S.~Khan, S.~Aaltonen, R.~Koblitz,
  M.~Holma, S.~Macchi, M.~E. Ruggieri \emph{et~al.}, ``Energy savings under
  performance constraints via carrier shutdown with {B}ayesian learning,'' in
  \emph{Proc. EuCNC}, 2023, pp. 1--6.

\bibitem{tekgul2023joint}
E.~Tekgul, T.~Novlan, S.~Akoum, and J.~G. Andrews, ``Joint uplink-downlink
  capacity and coverage optimization via site-specific learning of antenna
  settings,'' \emph{IEEE Trans. Wireless Commun.}, 2023.

\bibitem{de2023towards}
E.~de~Carvalho, A.~Valcarce, and G.~Geraci, ``Towards mobility management with
  multi-objective {B}ayesian optimization,'' in \emph{Proc. IEEE WCNC}, 2023,
  pp. 1--6.

\bibitem{frazier2018tutorial}
P.~I. Frazier, ``A tutorial on {B}ayesian optimization,''
  \emph{arXiv:1807.02811}, 2018.

\bibitem{3GPP38.843}
{3GPP Technical Report 38.843}, ``Study on artificial intelligence
  ({AI})/machine learning ({ML}) for {NR} air interface ({R}elease 18),'' Jun.
  2023.

\bibitem{3GPP38901}
{3GPP Technical Report 38.901}, ``Study on channel model for frequencies from
  0.5 to 100 {GHz} ({R}elease 16),'' Dec. 2019.

\bibitem{kelly1998rate}
F.~P. Kelly, A.~K. Maulloo, and D.~K.~H. Tan, ``Rate control for communication
  networks: {S}hadow prices, proportional fairness and stability,''
  \emph{Journal of the Operational Research society}, vol.~49, pp. 237--252,
  1998.

\bibitem{ye2013user}
Q.~Ye, B.~Rong, Y.~Chen, M.~Al-Shalash, C.~Caramanis, and J.~G. Andrews, ``User
  association for load balancing in heterogeneous cellular networks,''
  \emph{IEEE Trans. Wireless Commun.}, vol.~12, no.~6, pp. 2706--2716, 2013.

\bibitem{huang2006sequential}
D.~Huang, T.~T. Allen, W.~I. Notz, and R.~A. Miller, ``Sequential kriging
  optimization using multiple-fidelity evaluations,'' \emph{Structural and
  Multidisciplinary Optimization}, vol.~32, pp. 369--382, 2006.

\bibitem{balandat2020botorch}
M.~Balandat, B.~Karrer, D.~Jiang, S.~Daulton, B.~Letham, A.~G. Wilson, and
  E.~Bakshy, ``Botorch: {A} framework for efficient {M}onte-carlo {B}ayesian
  optimization,'' \emph{Advances in Neural Information Processing Systems},
  vol.~33, pp. 21\,524--21\,538, 2020.

\bibitem{eriksson2021high}
D.~Eriksson and M.~Jankowiak, ``High-dimensional {B}ayesian optimization with
  sparse axis-aligned subspaces,'' in \emph{Uncertainty in Artificial
  Intelligence}.\hskip 1em plus 0.5em minus 0.4em\relax PMLR, 2021, pp.
  493--503.

\bibitem{shen2021computationally}
Y.~Shen and C.~Kingsford, ``Computationally efficient high-dimensional
  {B}ayesian optimization via variable selection,'' \emph{arXiv:2109.09264},
  2021.

\bibitem{eriksson2019scalable}
D.~Eriksson, M.~Pearce, J.~Gardner, R.~D. Turner, and M.~Poloczek, ``{Scalable
  global optimization via local {B}ayesian optimization},'' \emph{Advances in
  neural information processing systems}, vol.~32, 2019.

\bibitem{lin2023overview}
X.~Lin, ``An overview of the {3GPP} study on artificial intelligence for {5G}
  new radio,'' \emph{arXiv:2308.05315}, 2023.

\bibitem{pan2023joint}
H.~Pan, Y.~Liu, G.~Sun, J.~Fan, S.~Liang, and C.~Yuen, ``Joint power and 3d
  trajectory optimization for uav-enabled wireless powered communication
  networks with obstacles,'' \emph{IEEE Trans. on Comm.}, vol.~71, no.~4, pp.
  2364--2380, 2023.

\bibitem{daulton2022multi}
S.~Daulton, D.~Eriksson, M.~Balandat, and E.~Bakshy, ``Multi-objective
  {B}ayesian optimization over high-dimensional search spaces,'' in
  \emph{Uncertainty in Artificial Intelligence}.\hskip 1em plus 0.5em minus
  0.4em\relax PMLR, 2022, pp. 507--517.

\bibitem{yang2019multi}
K.~Yang, M.~Emmerich, A.~Deutz, and T.~B{\"a}ck, ``Multi-objective {B}ayesian
  global optimization using expected hypervolume improvement gradient,''
  \emph{Swarm and evolutionary computation}, vol.~44, pp. 945--956, 2019.

\bibitem{emmerich2006single}
M.~T. Emmerich, K.~C. Giannakoglou, and B.~Naujoks, ``Single-and multiobjective
  evolutionary optimization assisted by {G}aussian random field metamodels,''
  \emph{IEEE Trans. on Evolutionary Computation}, vol.~10, no.~4, pp. 421--439,
  2006.

\bibitem{daulton2021parallel}
S.~Daulton, M.~Balandat, and E.~Bakshy, ``Parallel {B}ayesian optimization of
  multiple noisy objectives with expected hypervolume improvement,''
  \emph{Advances in Neural Information Processing Systems}, vol.~34, pp.
  2187--2200, 2021.

\bibitem{benzaghta2025wcnc}
M.~Benzaghta, G.~Geraci, D.~L{\'o}pez-P{\'e}rez, and A.~Valcarce, ``Data-driven
  optimization and transfer learning for cellular network antenna
  configurations,'' in \emph{Proc. IEEE WCNC}, 2025, pp. 1--6.

\bibitem{hoydis2023sionna}
J.~Hoydis, F.~A. Aoudia, S.~Cammerer, M.~Nimier-David, N.~Binder, G.~Marcus,
  and A.~Keller, ``Sionna {RT}: {D}ifferentiable ray tracing for radio
  propagation modeling,'' \emph{arXiv:2303.11103}, 2023.

\bibitem{LopPioGer2024}
D.~Lopez-Perez, N.~Piovesan, and G.~Geraci, ``Capacity and power consumption of
  multi-layer {6G} networks using the upper mid-band,''
  \emph{arXiv:2411.09660}, 2024.

\bibitem{benzaghta2025pimrc}
M.~Benzaghta, S.~Ammar, D.~L{\'o}pez-P{\'e}rez, B.~Shihada, and G.~Geraci,
  ``Data-driven design of {3GPP} handover parameters with {Bayesian}
  optimization and transfer learning,'' \emph{arXiv:2504.02633}, 2025.

\end{thebibliography}

\end{document}